\documentclass{ecai}  

\usepackage{graphicx}
\usepackage{latexsym}

\usepackage{balance} 

\usepackage{booktabs} 
\usepackage[ruled]{algorithm2e} 
\usepackage{multirow}
\usepackage{multicol}
\usepackage{bm} 
\usepackage{stfloats}

\usepackage[figuresright]{rotating}

\usepackage{perpage}
\MakePerPage{footnote}

\usepackage{enumitem}
\usepackage{float}
\usepackage{booktabs}

\usepackage{diagbox}
\usepackage{graphicx}

\usepackage{changes}
\usepackage{enumitem}
\usepackage{float}
\usepackage{array}
\usepackage{subfigure}
\usepackage[bottom]{footmisc}

\usepackage{caption}
\usepackage{hyperref}
\usepackage{amsfonts,amssymb, amsmath} 

\newtheorem{definition}{Definition}

\begin{document}

\begin{frontmatter}

\title{Learning to Collaborate by Grouping: a Consensus-oriented Strategy for Multi-agent Reinforcement Learning}

\author[A]{\fnms{Jingqing}~\snm{Ruan}}
\author[B]{\fnms{Xiaotian}~\snm{Hao}}
\author[C]{\fnms{Dong}~\snm{Li}\thanks{Corresponding Author. Email: lidong106@huawei.com}} 
\author[D]{\fnms{Hangyu}~\snm{Mao}}

\address[A]{Institute of Automation, Chinese Academy of Sciences}
\address[B]{Tianjin University}
\address[C]{Noah’s Ark Lab, Huawei}
\address[D]{SenseTime Research}

\begin{abstract}
Multi-agent systems require effective coordination between groups and individuals to achieve common goals. However, current multi-agent reinforcement learning (MARL) methods primarily focus on improving individual policies and do not adequately address group-level policies, which leads to weak cooperation. 
To address this issue, we propose a novel {\it Consensus-oriented Strategy} (CoS) that emphasizes group and individual policies simultaneously. 
Specifically, CoS comprises two main components: (a) the vector quantized group consensus module, which extracts discrete latent embeddings that represent the stable and discriminative group consensus, and (b) the group consensus-oriented strategy, which integrates the group policy using a hypernet and the individual policies using the group consensus, thereby promoting coordination at both the group and individual levels. Through empirical experiments on cooperative navigation tasks with both discrete and continuous spaces, as well as google research football, we demonstrate that CoS outperforms state-of-the-art MARL algorithms and achieves better collaboration, thus providing a promising solution for achieving effective coordination in multi-agent systems.
\end{abstract}

\end{frontmatter}

\section{Introduction}
%

Many applications, such as multi-player games~\cite{kurach2020google}, 
traffic signal control~\cite{wu2020multi},
and sensor networks~\cite{zhang2011coordinated},
can be modeled as cooperative multi-agent systems (MASs), where a team of agents performs a shared task to reach a common goal. A promising solution is cooperative multi-agent reinforcement learning (MARL) which has shown exceptional results, of which the popular approach for multi-agent cooperation is communication-based MARL. One line of research lies on fine-grained communication channels, such as graph neural networks~\cite{jiang2018graph,niu2021multi}, 
attention mechanisms~\cite{liu2022self,jiang2018atoc}, etc. Another line~\cite{das2019tarmac,du2021flowcomm,liu2020when2com} 
yields agents that learn communication protocols to determine which messages to transmit, and who to communicate with to assist decision-making.
However, a common issue is a lack of reasonable labor division for multiple agents.

In order to achieve better collaboration, many realistic multi-agent problems require a reasonable division of labor to enhance team-level cooperation~\cite{ferber1998meta,alam2016multi}. 
For example, in search and rescue missions, multiple agents (e.g. drones, robots, doctors) need to coordinate their efforts to locate and rescue survivors in a disaster-stricken area. Multiple drones should be deployed to search the area to ensure complete coverage of the search area and avoid collisions, while the doctors and robots should be assigned dynamically to different areas to ensure efficient rescuing. Thus, a well-designed cooperative multi-agent system can dynamically group agents with similar abilities based on the specific situation, while also allowing individual agents to make rational decisions based on their specific observations. 

Some methods have been proposed from the perspective of roles or groups. ROMA~\cite{wang2020roma} and RODE~\cite{wang2021rode} focus on task decomposition and specialize the agent associated with a role to resolve a certain sub-task. LDSA~\cite{yang2022ldsa} learns dynamic subtask assignments, which can dynamically group agents with similar abilities into the same subtask. However, on the one hand, role (group) representations are not well constrained and vulnerable to dynamic changes from the environment or policy training. On the other hand, the above methods rarely consider extracting higher-level policy guidance from the natural properties of groups (sub-tasks). In summary, good teamwork often requires a good division of labor, forward-looking guidance to specialize in a certain subtask, and rational individual decision-making. This poses a challenge for MARL for providing stable and powerful group representations, higher guidance at the team level, and better decisions at the individual level.

Therefore, we propose CoS, a consensus-oriented strategy in MARL. 
First, we use the vector quantized variational autoencoder (VQ-VAE)~\cite{van2017vqvae} to extract the group consensus embedding, which captures the shared objective that agents in the same group should pursue and  is essential for promoting effective teamwork.
Then, we perform the policy learning from two levels. At the higher level, we propose using a hyper-network architecture~\cite{ha2016hypernetworks}
to transform group consensus embedding into group-level decisions from global and long-term perspectives. At the lower level, we utilize the group consensus embedding as the context prompt to augment the observation to make the individual and refined policy. The combination of these two policies guarantees the foresight and precision of the decision-making process. Moreover, CoS is a pluggable module and is suitable for both discrete and continuous action spaces.
We evaluate our method in three challenging MARL environments including discrete cooperative navigation, continuous cooperative navigation~\cite{lowe2017multi}, and google research football~\cite{kurach2020google}. The results show that our CoS significantly improves the learning performance on these benchmarks compared to some competitive baselines.



\section{Related Work}

\paragraph{\textbf{Multi-agent Grouping}.}
We classify existing algorithms into three categories: (i) prior knowledge-based, (ii) role-based, and (iii) subtask-based methods.
Some early methods~\cite{odell2002role,lhaksmana2013role,ferber2004agents,pavon2003agent} utilize the domain knowledge to group the agents, and are limited by human labor.
To solve the issue, one line of work groups multiple agents by generating diverse roles responsible for different parts. 
ROMA~\cite{wang2020roma} learns a role-specific policy where the roles are captured from agents' local observations.
ROGC~\cite{liu2022rogc} introduces the graph convolutional network for classifying agents into different roles.
RODE~\cite{wang2021rode} decomposes action spaces with the learned roles.
\cite{hu2022policy}~define the role and role diversity to measure a cooperative MARL task and help diagnose the current policy.
LILAC~\cite{fu2022lilac} learns a leader to assign roles.
Another line of work such as~\cite{yang2022ldsa,yuan2022multi,phan2021vast,iqbal2021randomized}, divides the agents into some groups that carry out similar sub-tasks with a specific policy or value function.
In our work, we learn more stable and distinguishable group embeddings and further consider the integration of team-level strategy and individual-level decision.

\vspace{-6pt}

\paragraph{\textbf{Representation Learning in MARL}.}
The representations for observations, actions, or underlying messages are widely studied in MARL.
Some works ~\cite{aman2021knowledge,fervari2022bisimulations} use bisimulation metrics to extract the latent embeddings from observations.
\cite{li2022ace,agarwal2021contrastive} attempt to learn  action representations to assist multi-agent policy learning. 
\cite{xu2022consensus,jiang2022multi} propose represent underlying messages to conduct effective communication in MAS.
There are also some works like~\cite{tianhypertron,zhang2022common} that use VAE to encode the trajectory message to make representation more knowledgeable.
Unlike the above methods, we introduce VQ-VAE to maintain a quantized hidden space to extract stable,  and distinguishable group consensus embeddings to associate a more powerful strategy.




\section{Preliminary}

\paragraph{\textbf{Vector Quantised-Variational AutoEncoder}.}
The VQ-VAE is a type of variational autoencoder that uses vector quantization to obtain a discrete latent representation. 
It includes an encoder $z_e$, a decoder $z_d$, and a codebook $e \in \mathbb{R}^{K \times D}$, where $K$ is the number of embeddings in the codebook and $D$ is the dimension of the embeddings.
The encoder maps the input data $x$ to a sequence $z$ of discrete codes from a codebook and the decoder takes the responsibility of reconstruction.
In our paper, we use the encodings in the codebook as stable and distinguishable representations.

Formally, $z_e(x)$ is mapped via nearest-neighbor into the quantized codebook, denoted as:
\begin{equation}
    z_q(x) = e^j \ where \ j = argmin_k || z_e(x) - e^k||_2^2.
\end{equation}

This discretized process is not differentiable, thus copying the gradient of $z_q(x)$ to $z_e(x)$ is a suitable approximation similar to the straight-through estimator. The loss can be written as follows.
\begin{equation}
    \mathcal{L} = ||z_d(z_q(x))-x||^2_2 + \beta ||sg[z_e(x)]-e^j||^2_2 + ||z_e(x),sg[e^j]||^2_2,
\end{equation}
where $sg$ is a stop gradient operator, and $\beta$ is a parameter that regulates the rate of code change.


\section{Methodology}


\subsection{Problem Setup}
\label{prob-setup}
\paragraph{\textbf{Group Consensus-guided Dec-POMDP.}}
We formalize our problem with multiple agents as
a group consensus-guided decentralized partially observable Markov decision process (GC-Dec-POMDP). A GC-Dec-POMDP is given by a tuple $(\mathcal{I}, \mathcal{S}, \mathcal{O}, \mathcal{G}, \mathcal{A}, P, R, \gamma)$. 
$\mathcal{I}$ denotes the set of agents indexed by $i \in {1,...,N}$.
Let $\mathcal{S}$ represent the global state space and $\mathcal{A}$ be the action space of the agent. $\mathcal{G}={\{e_1, e_2, ..., e_{K}\}}$ denotes the group codebook to generate the group consensus embedding $e^j$ belonging to the group $j$ for each agent $i$ in every time step, which divides all the agents into various parts to realize the corresponding density of coordination. 
At each time step $t$, each agent $i$ chooses its action $a^i_t \in \mathcal{A}$ based on the group consensus policy $\pi^{{gc}^j_i}$ conditioned only on the group consensus embedding $e^j$ and the group guided policy $\pi^{{gg}^j_i}$ conditioned on $e^j$ and its observation drawn from the observation function $O(s_t, i)$ where $s_t \in \mathcal{S}$, and receive the environmental reward $r_t$ given by its reward function $R^i:\mathcal{O} \times \mathcal{A} \to \mathbb{R}$.
$P(s_{t+1}|s_t,\bm{a}_t)$ is the dynamics function that gives the distribution of the next state $s_{t+1}$ at the current state $s_t$ to execute the joint action $\bm{a}_t=\{a^i_t\}_{i=1}^N$.
In summary, each agent learns the group consensus policy $\pi^{{gc}^i}$ and the group guided policy $\pi^{gg^i}$ simultaneously, denoted as $\pi^i=(\pi^{{gc}^i}, \pi^{{gg}^i})$
, and the overall objective is to find the optimal joint policy $\bm{\pi}=(\pi_1,...,\pi_N)$ such that the discounted returns of each agent $G_i = {\sum\limits_{t = 1}^\infty  {{\gamma ^{t-1}}r_t^i}}$ are maximized, where $\gamma \in [0,1)$ is a discounted factor. 

\paragraph{\textbf{Optimization Target.}}
The overall goal is to maximize the cumulative return, denoted as:
\begin{equation}
\label{eq:ret}
\eta  = {\mathbb{E}_{\bm{a} \sim \bm{\pi} }}\left[ {\sum\limits_{k = 0}^\infty  {{\gamma ^k}} r({s_{t + k}},{\bm{a}_{t + k}})} \right],
\end{equation}
where $\bm{\pi}$ is the joint policy for all the agents, formulated as:
\begin{equation}
    \bm{\pi}(\bm{a}|\{o^i\}_{i=1}^N) \buildrel \Delta \over = \{\pi^{{gc}^i}(a^{gc^i_j} | e^j)  +  \pi^{{gg}^i}(a^{gg^i_j}|o^i,e^j) \}_{i=1}^N
\end{equation}
where $e^j$ is the group consensus embedding. Here, we drop the time step $t$ for simplification.

\begin{figure*}[htbp]
    \centering
    \includegraphics[width=0.8\linewidth]{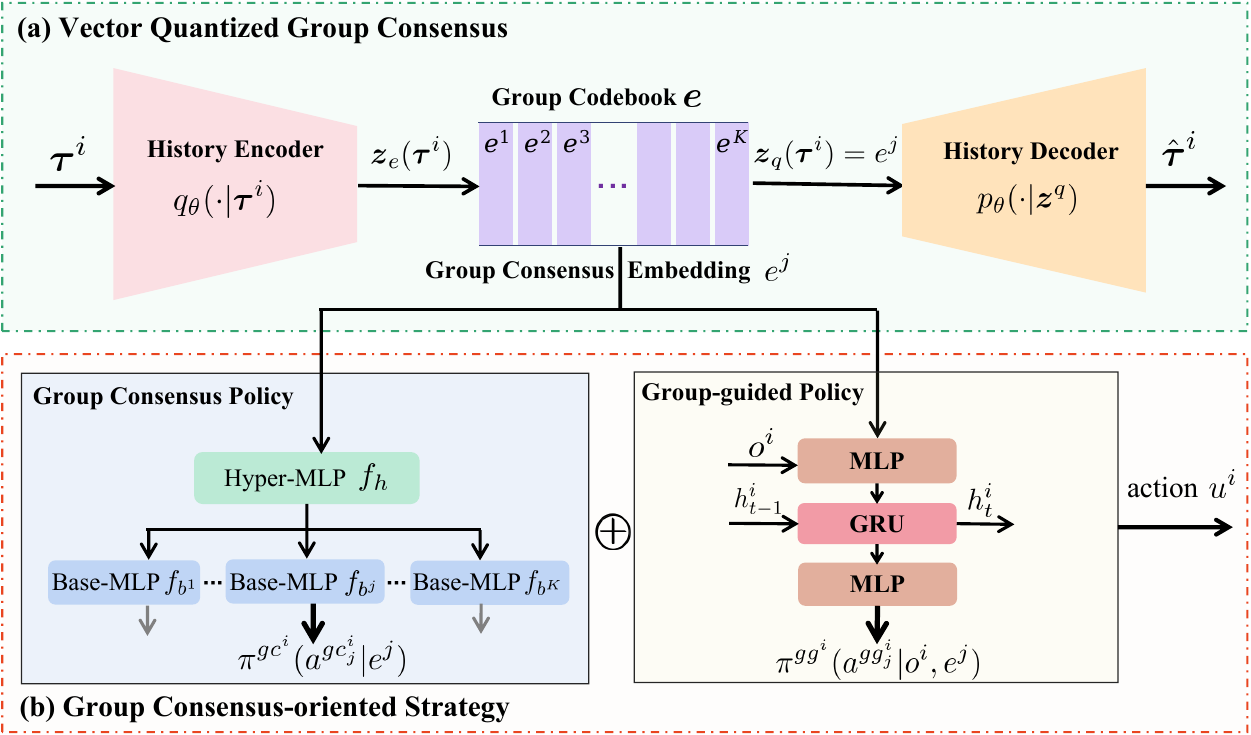}
    \caption{The schematics of our framework CoS. CoS includes two parts: (a) The extraction of group consensus embeddings is finished by the module of vector quantized group consensus. (b) The group consensus-oriented strategy is utilized to generate the overall decisions.}
    \label{fig:main}
\end{figure*}

\subsection{The Motivation and Framework}
\label{subsec:framework}

Effective collaboration is a crucial aspect of multi-agent systems and relies heavily on well-coordinated teamwork. One approach to achieving effective collaboration is through the use of consensual groups, which consist of one or more agents guided by a shared pattern to make higher-level decisions for completing a particular sub-task. To ensure the success of the consensual groups, we consider three fundamental principles:

a) \textbf{Group Consensus}: Agents within a group should strive to reach a consensus to accomplish a specific sub-task. The group consensus should be well-defined and distinguishable, with sufficient knowledge.

b) \textbf{Behavioral Diversity}: Both inter- and intra-group agents should exhibit diverse and varied behavioral patterns to promote exploration and better solutions.

c) \textbf{Dynamic Integration}: The multi-agent decision-making process should involve a dynamic combination of a group consensus policy and an individual policy guided by group knowledge. This approach will allow for flexibility in adapting to changing circumstances and enable the agents to make informed decisions in a collaborative manner.






Taking the above factors into consideration, we present the \textbf{C}onsensus-\textbf{o}riented \textbf{S}trategy (CoS) learning framework, shown in Figure~\ref{fig:main}.
The upper part is designed for extracting knowledgeable and distinguishable group consensus embedding. The lower part illustrates the group consensus-oriented strategy. 
Compare to common MARL methods, we innovatively introduce the VQ-VAE for learning group consensus embedding and propose an additional group consensus policy. This embedding module can be considered as observation augmentation, and the group consensus policy can be incorporated into most MARL algorithms to complement their policy.
Moreover, the network architecture and the detailed parameter settings can be found in Appendix~\ref{app:net} and~\ref{app:para}, respectively. The pseudo-code for CoS is provided in~\ref{app:algo}.

\subsection{Vector Quantized Group Consensus}
\label{subsec:vqgc}

Drawing inspiration from the vector-quantized variational autoencoder (VQ-VAE) proposed by~\cite{van2017vqvae}, we introduce a novel module called Vector Quantized Group Consensus (VQGC) to achieve identifiable and rich group consensus. The motivations behind this include: 
a) The discrete codebook in VQ-VAE can effectively group features with similar semantics into tighter clusters with better separation than traditional continuous representation methods. This enables more robust and efficient feature extraction in multi-agent settings.
b) The range of VQ encodings is highly controllable, and training the VQ encodings as group embeddings of agents leads to reduced variance, thus enhancing the stability of multi-agent collaboration.
c) The use of a discrete feature representation can mitigate the effects of semantic noise in stochastic environments by limiting the number of possible bias vectors. This, in turn, ensures the extraction of richer knowledge from the features.


Thus, building on the powerful properties of VQ-VAE, we design the VQGC to extract the group consensus embeddings,  which enables neural models to learn similarities and differences between states better and generate a higher-level consensual representation to guide policy learning. Next, the design details are elaborated as follows.

\paragraph{\textbf{History Encoder}.}
At time step $t$, given the history transitions $\tau^i$ including observations $o^i$, action $a^i$, and reward $r^i$, we first aim to extract the feature by a history encoder parameterized by $\theta$, denoted as:
\begin{equation}
    z_e(\bm{\tau}^i) = q_\theta(\cdot|\bm{\tau}^i),
\end{equation}
where $\bm{\tau}^i:=\{ o^i_l, a^i_{l-1}, r^i_{l-1} \}_{l=t-c}^t$ and $c$ is the window size of the history chunk.

\paragraph{\textbf{Group Codebook}.}
The group codebook refers to a set of vectors used in the VQGC module. 
Specifically, it is defined as $\bm{e} = \{e^1,e^2,...,e^K\}, \forall e^j \in \mathbb{R}^{1 \times D}$, where $j$ is the $j^{th}$ entry and $D$ is the dimensionality of each entry.  
To obtain a powerful and robust group codebook, we introduce a  regularizer inspired by~\cite{fan2021multi} into the training process.
Euclidean space has limited representation capacity in a fixed dimension, and increasing the dimensionality will bring a large computation budget and training overfitting issues. Thus, we consider a hyperbolic space constraint to generate more powerful and knowledgeable embedding representations.
Thus, we give a definition of the Poincar\'e ball model~\cite{poincare1882theorie}.
\begin{definition}
The Poincar\'e ball model is a model of $n$-dimensional ($n \ge 3$) hyperbolic geometry in which the points of the geometry are in the $n$-dimensional unit ball.
\end{definition}

Let a Poincar\'e ball with dimension $d$ and radius $1$ be $\mathcal{P}^{d,1}:=\{\bm{e} \in \mathbb{R}^d, ||\bm{e}|| < 1\}$, and the operator $||\cdot||$ denotes the Euclidean $L^2$ norm. 
This corresponds to the Riemannian manifold $(\mathcal{P}^{d,1}, g_{\bm{e}})$, where $g_{\bm{e}} = (2/(1-||\bm{e}||^2))^2 g^E$ is the Riemannian metric tensor and $g^E$ denotes the Euclidean metric tensor. 
Then the distance between two vector $\bm{x}, \bm{y} \in \mathcal{P}^{d,1}$ can be computed as:
\begin{equation}
    d(\bm{x},\bm{y})=arcosh\left ( 1+2\frac{||\bm{x}-\bm{y}||^2}{(1-||\bm{x}||^2)(1-||\bm{y}||^2)}  \right ) .
\end{equation}

For inducing an appropriate structural bias on the group embedding space, we introduce a novel metric $\mathcal{L}_{\mathcal{P}}$ to constrain the embeddings into hyperbolic space with the Poincar\'e ball model, which is well-suited for the gradient-based optimization~\cite{nickel2017poincare}, formulated as:
\begin{equation}
\mathcal{L}_{\mathcal{P}}=\sum_{(\bm{e}^I,e^j)\in \mathcal{B}_e}  log\sigma( Sgn((\bm{e}^I,e^j)) \cdot D(\bm{e}^I,e^j))  ,
\end{equation}
where $D(\bm{e}^I,e^j) = \min_{i \in I} d(e^i, e^j)$ is the shortest Poincar\'e distance for the anchor $e^j$ and other entries $\bm{e}^I$ sampled from the buffer $\mathcal{B}_e$ that is used to save the latest $L$ group codebooks. This buffer at time step $t$ is denoted as $\mathcal{B}_t^e = \{e^1_l,e^2_l,...,e^K_l\}_{l=t-L}^t$.  $\sigma$ is a logistic activation function and $Sgn$ is a symbolic function defined as follows.
\begin{equation}
    Sgn((\bm{e}^I,e^j))=\left\{\begin{matrix}
        1, \ \ \ \ if \ \bm{e}^I \in Pos
         \\
        -1, \ \ \ if \ \bm{e}^I \in Neg
    \end{matrix}\right.
\end{equation}
where the set $Pos$ denotes the identical group embedding $e^i_l$ with different time steps. The set $Neg$ represents the different group embeddings $e^j_l$, where $j \ne i, \forall l \in (t-L,t]$.

By minimizing $\mathcal{L}_{\mathcal{P}}$, we can pull together the group embeddings with the same identifier to relieve the non-stationarity of drastic changes and push apart different group embeddings to obtain distinguishable representations.



Given the current group codebook $\bm{e}$ and the encoded feature $z_e(\bm{\tau}^i)$, the nearest matching mechanism can be denoted as $Quantize$, formulated as follows.
\begin{equation}
\label{eq:quantize}
    \begin{matrix}
    Quantize(z_e(\bm{\tau}^i)) = e^j , \ \  \
    j=argmin_{k} ||z_e(\bm{\tau}^i) - e^k|| ,
    \end{matrix}
\end{equation}
where $k$ is the index of the length of the group codebook. In the following, we denote Equation~(\ref{eq:quantize}) as $e^j$ to represent the nearest quantized vector for brevity.

\paragraph{\textbf{History Decoder}.}

The history decoder takes the $z_q(\bm{\tau}^i)$ as the input to reconstruct the input $\bm{\tau}^i$ that is to maximize the log-likelihood:
\begin{equation}
    \mathcal{L}_{recon}=\mathbb{E}_{z_q}\left [  log p(\hat \tau^i | z_q(\tau^i))  \right ].
\end{equation}
Thus, the decoding of the history transitions can capture the dynamics of the environment to a large extent.


 
\paragraph{\textbf{Training Objective}.}
Based on VQVAE, the training of VQGC is equipped with the stop gradient technique, summarized as follows.
\begin{equation}
\label{eq:vqgc}
    \begin{matrix}
    \mathcal{L}_{VQGC}=\underbrace{log p _\theta(\hat {\tau}^i | z_q({\tau}^i))}_{history \ decoder} + \beta \underbrace{||sg[z_e({\tau}^i)]-e^j||^2_2}_{history \ encoder}
     \\
    +  \underbrace{||z_e({\tau}^i),sg[e^j]||^2_2 + \mathcal{L}_{\mathcal{P}} }_{group \ codebook}
    \end{matrix},
\end{equation}
where the operator $sg$ indicates the stop of gradient backpropagation. 
$\beta$ is the hyper-parameter preventing the encoder outputs from fluctuating between different code vectors.

So far, the group consensus embedding is outputted by our VQGC. Through training, the group consensus embedding fully incorporates rich dynamical knowledge, as well as robust and distinguishable representations. 
Next, we will introduce the group consensus-oriented strategy to use this extracted knowledge to provide global intention and individual guidance for the multi-agent decision-making process.


\subsection{Group Consensus-oriented Strategy}
\label{subsec:gcs}

Our group consensus-oriented strategy includes two parts: group consensus policy aims to generate the higher-level decision from the global and long-term perspective and group-guided policy is used to make individual decisions condition on the group consensus, elaborated as follows.

\paragraph{\textbf{Group Consensus Policy (GCP)}.}

Here, we propose using a hyper-network architecture where a primary network $f_h$ generates the weights to  parameterize all the base layer $f_b^i$, where $ i \in \{1,2,..., K\}$. 
A hyper-network is a network that generates the weights for another network.
Specifically, VQGC generates the group consensus embedding $e^j$ for each agent $i$. Then the hyper-MLP takes the group consensus embeddings $e^j$ as input and produces the weights for each Base-MLP network which performs fine-grained control denoted as $f_{b^j}=\pi^{gc^i}(a^{gc_j^i}|e^j)$.
The hyper-MLP can be formulated as $f_h: \mathbb{R}^d \mapsto \mathbb{R}^{dim(b^j)}$ which is parameterized by $h$, $d$ and $dim(b^j)$ are the dimensions of group embedding and base network, respectively.
The formulation of a Base-MLP $f_{b^j}$ is given by $f_{b^j}(e^j) = \bm{W}_1(\sigma (\bm{W}^T_2 e^T_j))$, where the weight matrix $\bm{W}_1$ and $\bm{W}_2$ are produced by Hyper-MLP $f_h$.
 
The motivation for such a design centrally involved two aspects. On the one hand, agents from different groups are expected to solve different sub-tasks on different parameter spaces, which escalates the difficulties of policy learning. The hyper-network fed with different group consensus embeddings can well coordinate the potentially conflicting parameters in a unified space, which reduces the training complexity to some extent. On the other hand, the hyper-network captures higher-order interaction among group consensus embeddings, which is conducive to making global decisions.


\paragraph{\textbf{Group-Guided Policy (GGP)}.}

Besides the higher-level decision, the individual policy guided by the group that emphasizes the underlying dynamic of the environment is also essential. Here, we utilize the group consensus embedding $e^j$ as the context prompt to augment the observation $o^i$ for agent $i$ belonging to the group $j$, denoted as $\pi^{gg^i}(a^{gg^i_j}|o^i,e^j)$.
The group-guided policy is used to make individual decisions conditioned on the group consensus.


\paragraph{\textbf{Training Objective}.}
Borrow the policy gradient for perturbation network~\cite{fujimoto2019off}, the final executed action of the agent $i$ can be denoted as follows.
\begin{equation}
\label{eq:pi}
  u^i = \pi^{{gc}^i}( e^j)  +  \pi^{{gg}^i}(o^i,e^j)   
\end{equation}

With Equation~(\ref{eq:ret}) and (\ref{eq:pi}), the overall optimization objective for the expected cumulative return can be written as:
\begin{equation}
    \mathcal{J}=\mathbb{E}_{s,e^j,\bm{u}}\left [ \prod_{i=1}^{N}\left (  \pi^{gc^i}(e^j)+\pi^{gg^i}(o^i,e^j) \right ) \cdot Q_{\bm{\pi}}(s,\bm{u})  \right ] ,
\end{equation}
where $s\sim p^{\bm{\pi}}$ is the state sampled from the stationary distribution $p^{\pi}$, $e^j \sim z_q$ is the generated group embedding, and $\bm{u} \sim \{\pi^{gc^i}+\pi^{gg^i}\}_{i=1}^N$ denotes the joint action.

Thus, the gradient for the group consensus policy of agent $i$ parameterized by $\phi$ can be derived by applying the mini-batch technique to the off-policy training:
\begin{equation}
\label{eq:gcp}
    \bigtriangledown_{\phi}^i \mathcal{J} = \mathbb{E}
\left [ \bigtriangledown_{\bm{u'}}Q_{\bm{\pi}}(s,\bm{u}')|_{\bm{u}'=\{\pi^{gc^i}_{\phi}+\pi^{gg^i}_{\varphi}\}} 
\bigtriangledown_{\phi}log\pi^{gc^i}_{\phi}(e^j)
 \right ] .
\end{equation}

Also, the gradient for the group-guided policy is shown as:
\begin{equation}
\label{eq:ggp}
    \bigtriangledown_{\varphi}^i \mathcal{J} = \mathbb{E}
\left [ \bigtriangledown_{\bm{u'}}Q_{\bm{\pi}}(s,\bm{u}')|_{\bm{u}'=\{\pi^{gc^i}_{\phi}+\pi^{gg^i}_{\varphi}\}} 
\bigtriangledown_{\phi}log\pi^{gg^i}_{\varphi}(o^i,e^j)
 \right ] .
\end{equation}

\paragraph{\textbf{Jump-start Dynamic Integration}.}
However, evidence from supervised learning suggests hypernetwork performance is highly sensitive to input and initialization.
Thus, we perform a trick called Jump-start Dynamic Integration (JDI) to alleviate the issue inspired by~\cite{uchendu2022jump}.
Here, we define a hyper-parameter $jump \in [0,1]$ to represent a proportion factor of the full training horizon $T$, which indicates it should be changed while the current time step $t$ satisfies $t \ge jump * T$. The change can be formulated as:
\begin{equation}
    \pi^{finall} = \begin{cases}
    \pi^{gg^i}(o^i,e^j), \ \  if \  t < jump*T
     \\
    \pi^{gc^i}(e^j)+\pi^{gg^i}(o^i,e^j), \ \ else
    \end{cases}.
\end{equation}

In our case, we adopt this trick to stabilize the training process of our group consensus policy by warming up the training of the group codebook.

\setlength{\tabcolsep}{2.8pt}
\setlength{\abovecaptionskip}{0.5pt}
\setlength{\belowcaptionskip}{0.5pt}
\begin{table*}[!ht]
\vspace{-6pt}
    \centering
    \caption{End steps of various methods on d-CN.}
    \begin{tabular}{ccccccc}
    \toprule
      \#agents & CoS & MAPPO & VDN & QMIX & RODE & ROMA \\ \hline
     4 & \textbf{19.8} $\pm$ \scriptsize{1.1} & 21.2 $\pm$ \scriptsize{0.7} & 24.6 $\pm$ \scriptsize{0.4} & 24.3 $\pm$  \scriptsize{0.3} & 23.6 $\pm$ \scriptsize{1.2} & 24.8 $\pm$ \scriptsize{0.9} \\ 
     6 & \textbf{23.1} $\pm$ \scriptsize{0.9} & 23.2 $\pm$ \scriptsize{0.8} & 24.9 $\pm$ \scriptsize{0.6} & 24.7 $\pm$ \scriptsize{0.5} & 24.9 $\pm$ \scriptsize{0.8} & 24.9 $\pm$ \scriptsize{0.6} \\ 
    10 & 25 $\pm$ \scriptsize{0} & 25 $\pm$ \scriptsize{0} & 25 $\pm$ \scriptsize{0} & 25 $\pm$ \scriptsize{0} & 25 $\pm$ \scriptsize{0} & 25 $\pm$ \scriptsize{0} \\   
    \bottomrule
    \end{tabular}
    \label{tab:endstep-d-CN}
\end{table*}


\section{Experimental Evaluations}

\begin{figure}[htbp]
    \subfigure[ d-CN or c-CN]{
    \label{fig:dcn-ccn}
		\begin{minipage}[t]{0.23\textwidth}
			\centering
			\includegraphics[width=\textwidth]{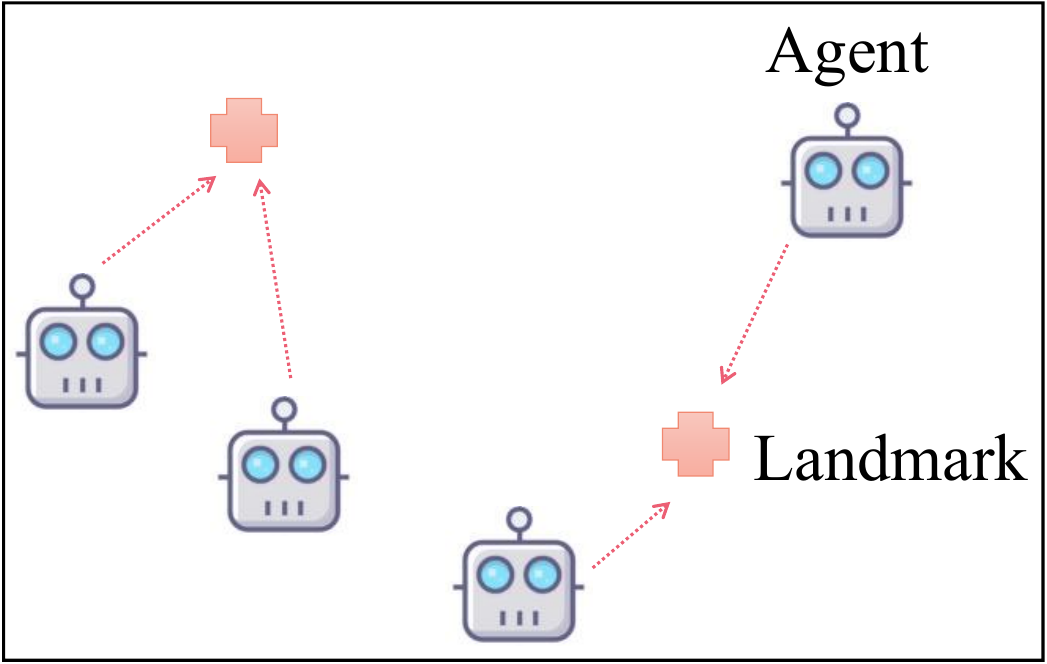}
		\end{minipage}
	}\vspace{-3pt}
	\subfigure[  GRF]{
 \label{fig:grf-}
		\begin{minipage}[t]{0.223\textwidth}
			\centering
			\includegraphics[width=\textwidth]{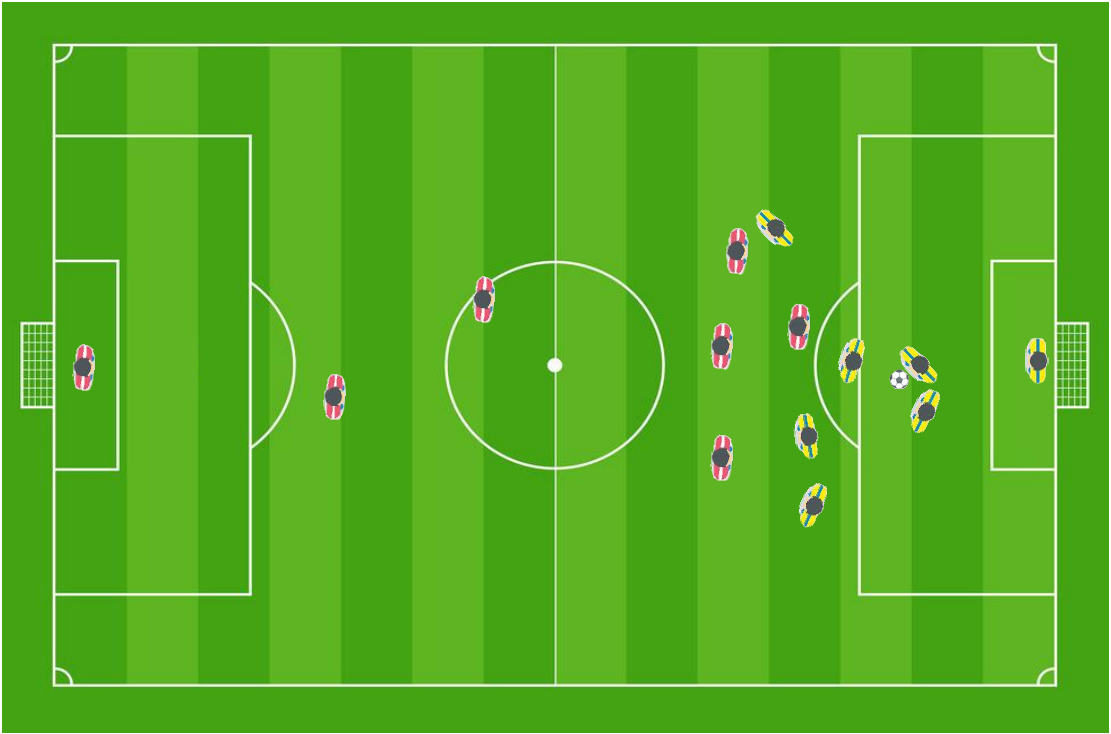}
		\end{minipage}
	}\vspace{-3pt}
	\centering
	\caption{The schematics of our experimental environments.}
 \vspace{-6pt}
	\label{fig:draw}
\end{figure}

We evaluate the effectiveness of our algorithm in three environments including both discrete and continuous action space: discrete Cooperative Navigation (d-CN), continuous Cooperative Navigation (c-CN), and Google Research Football (GRF), illustrated as Figure~\ref{fig:draw}.

\vspace{-10pt}
\vspace{-20pt}
\subsection{Experimental Settings}

\paragraph{\textbf{d-CN and c-CN}.}
We modify the classic Cooperative Navigation (CN) implemented in the multi-agent particle world~\cite{lowe2017multi} to a more challenging environment, which requires more collaboration among agents.
We initialize the CN world with $n$ landmarks and $2*n$ agents with random locations at the beginning of each episode.
Each agent can only observe its velocity, position, and displacement from other agents and landmarks. 
The final objective is to occupy all the landmarks and each landmark contains two agents.
The reward function can be formulated as $r_t = -0.1 + 3*single + 10*double$. The variables $single$ and $double$ denote the number of landmarks that are occupied by only one and two agents, which corresponds to the reward $3$ and $10$, respectively. 
$-0.1$ is the step punishment.
Obviously, the game can reach the maximum reward while each landmark contains two agents, and can be over. 
d-CN denotes that the world has a discrete action space $\mathcal{A}_d$ including five actions \textit{[up, down, left, right, stop]}.
c-CN represents that the world has a continuous action space $\mathcal{A}_c$.
We set the length of each episode as $25$ time steps.

\paragraph{\textbf{GRF}.}
GRF~\cite{kurach2020google} is a realistically complicated and dynamic multi-agent testbed. 
Agents should have a division of labor and plan to coordinate the time and location to complete the scoring.
In the experiments, we control left-side players except for the goalkeeper while the right-side players are built-in
bots controlled by the game engine. 
Here, each player has 19 actions to control,
including the standard move actions and different ball-kicking techniques.
The observation contains the positions and moving directions of the ego-agent, other agents, and the ball.
We use the Floats wrappers to represent the state that contains a 115-dimensional vector.
The rewards include the SCORING reward $\{-1, +1\}$, and
the CHECKPOINT reward, which is the shaped reward that specifically addresses the sparsity of SCORING. 
The detailed descriptions of GRF can be found in Appendix~\ref{app:grf}.

\paragraph{\textbf{Baselines}.}
We compared our results with several baselines as
follows. {VDN} and {QMIX} are state-of-the-art (SOTA) value factorization approaches, with which it is difficult to obtain coordinated behaviors.
{MAPPO} is the multi-agent competitive SOTA algorithm extended from PPO by setting the sharing actor and a centralized critic. 
{MADDPG} and {MATD3} are classic continuous control algorithms.
{ROMA} and {RODE} are role-based grouping algorithms.
Note that VDN, QMIX, ROMA, and RODE are designed for the discrete action space, while MADDPG and MATD3 are designed for the continuous action space.

\subsection{Does CoS Perform Better?}

For validating our CoS, we conduct empirical experiments on d-CN, c-CN, and GRF.
The benchmarks we choose include discrete and continuous action spaces, realistically complicated and stochastic worlds.
All experiments are repeated for 10 runs with different random seeds.


\begin{figure*}[ht]
    \centering
    \subfigure{
		\begin{minipage}[t]{0.4\textwidth}
			\centering
			\includegraphics[width=\textwidth]{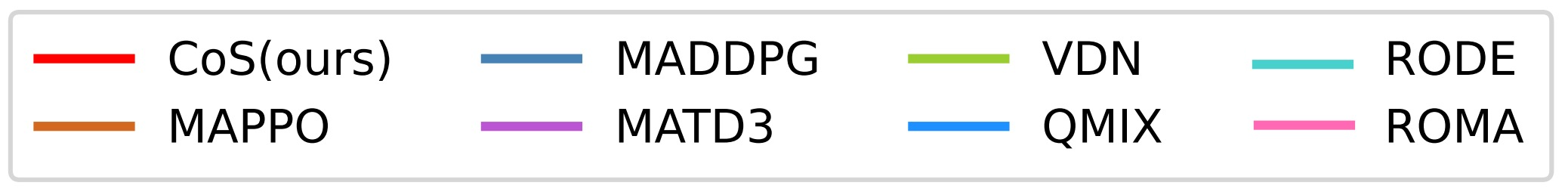}
		\vspace{-15pt}
		\end{minipage}
	}\vspace{-5pt}   
 
    \setcounter{subfigure}{0}
    \subfigure[\small 4 agents]{
    \label{fig:d-cn4}
		\begin{minipage}[t]{0.3\textwidth}
			\centering
			\includegraphics[width=\textwidth]{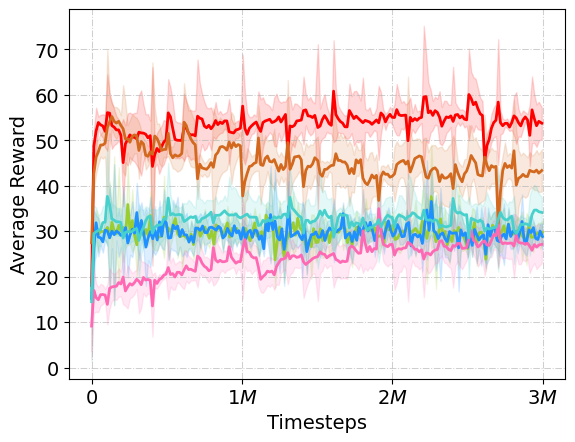}
		\end{minipage}
	}\vspace{-3pt}
	\subfigure[\small 6 agents]{
 \label{fig:d-cn6}
		\begin{minipage}[t]{0.3\textwidth}
			\centering
			\includegraphics[width=\textwidth]{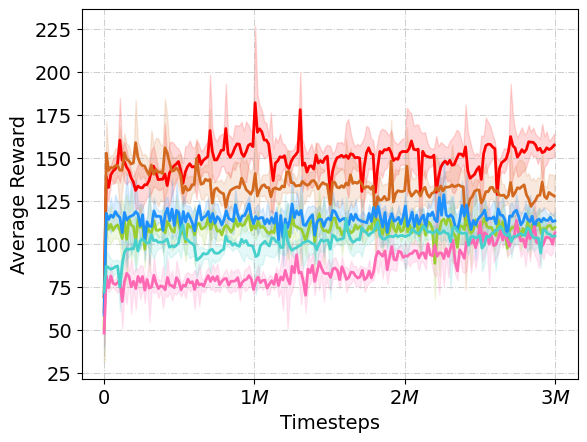}
		\end{minipage}
	}\vspace{-3pt} 
 \subfigure[\small 10 agents]{
 \label{fig:d-cn10}
		\begin{minipage}[t]{0.3\textwidth}
			\centering
			\includegraphics[width=\textwidth]{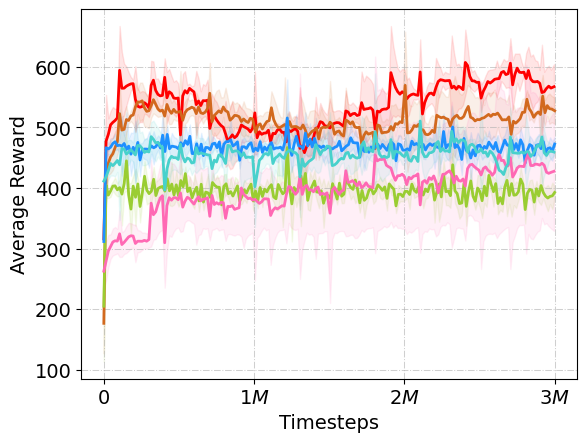}
		\end{minipage}
	}\vspace{-3pt} 
 
	\centering
 \subfigure[\small 4 agents]{
    \label{fig:c-cn4}
		\begin{minipage}[t]{0.3\textwidth}
			\centering
			\includegraphics[width=\textwidth]{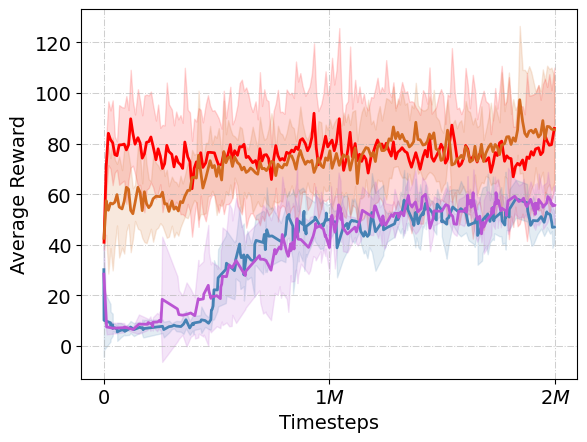}
		\end{minipage}
	}
	\subfigure[\small 6 agents]{
 \label{fig:c-cn6}
		\begin{minipage}[t]{0.3\textwidth}
			\centering
			\includegraphics[width=\textwidth]{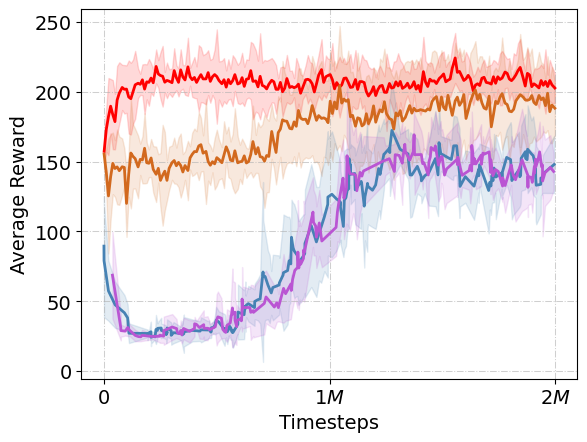}
		\end{minipage}
	}
 \subfigure[\small 10 agents]{
 \label{fig:c-cn10}
		\begin{minipage}[t]{0.3\textwidth}
			\centering
			\includegraphics[width=\textwidth]{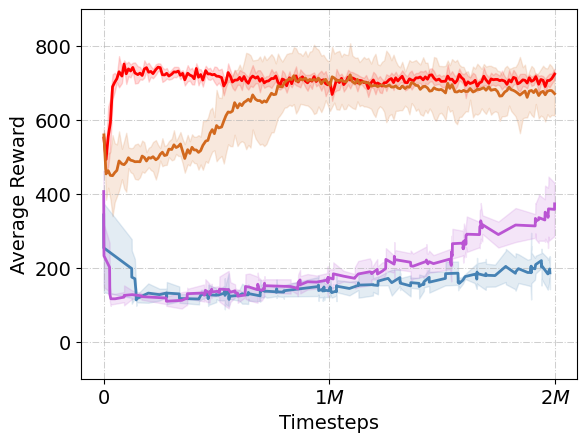}
		\end{minipage}
	}
	\centering
	\caption{Average episodic rewards and the confidence level for 4, 6, and 10 agents on CN. (a-c) The results on d-CN. (d-f) The results on c-CN. }
	\label{fig:exp-cn}
\end{figure*}

\subsubsection{Performance on d-CN.}
We conduct the experiments across $4,6$ and $10$ agents in d-CN with discrete action space as shown in Figure~\ref{fig:exp-cn}(a-c). 
CoS substantially gets a better average reward than all the baselines, indicating that the group consensus strategy increasingly enhances the superiority 
of our method.

Moreover, as shown in Table~\ref{tab:endstep-d-CN}, we report the mean end steps of these methods after testing 100 episodes, and the episode limit is set to 25. 
CoS basically completes the task faster with the least number of steps.
These results show that the group consensus strategy of Cos can group agents with different intentions toward various landmarks, which boosts the training performance.

\subsubsection{Performance on c-CN.}
As shown in Figure~\ref{fig:exp-cn}(d-f), the results in continuous space also exhibit better performance. 
Obviously, our method has a quicker convergence speed and a smaller variance than others, which indicates the decision of CoS to consider both global and individual guidance can further exploit the underlying properties and facilitate cooperative behaviors among agents.

In Table~\ref{tab:endstep-c-CN}, our method still get the least number of steps. Unfortunately, all the methods have failed with 10 agents because of the stochasticity and difficulty of the environment. It will be an interesting direction to study how to obtain the optimal solution in such complicated scenarios in the fastest time steps.

\setlength{\tabcolsep}{5.8pt}
\setlength{\abovecaptionskip}{0.5pt}
\setlength{\belowcaptionskip}{0.5pt}
\begin{table}[!ht]
    \centering
    \caption{End steps of various methods on c-CN.}
    \begin{tabular}{ccccccc}
    \toprule
      \#agents & CoS & MAPPO & MADDPG & MATD3 \\ \hline
    4 & \textbf{19.6} $\pm$ 2.1  & 20.3 $\pm$ 2.3  & 22.3 $\pm$ 1.2  & 22.1 $\pm$ 1.6   \\ 
    6 & \textbf{22.1} $\pm$ 1.4  & 23.8 $\pm$ 1.5   & 24.3  $\pm$ 1.3 & 24.2 $\pm$ 1.3   \\ 
    10 & 25 $\pm$ 0  & 25 $\pm$ 0   & 25 $\pm$ 0  & 25 $\pm$ 0   \\ 
\bottomrule
    \end{tabular}
    \label{tab:endstep-c-CN}
\end{table}


\begin{figure}[h!]
    \centering
    \subfigure{
		\begin{minipage}[t]{0.3\textwidth}
			\centering
			\includegraphics[width=\textwidth]{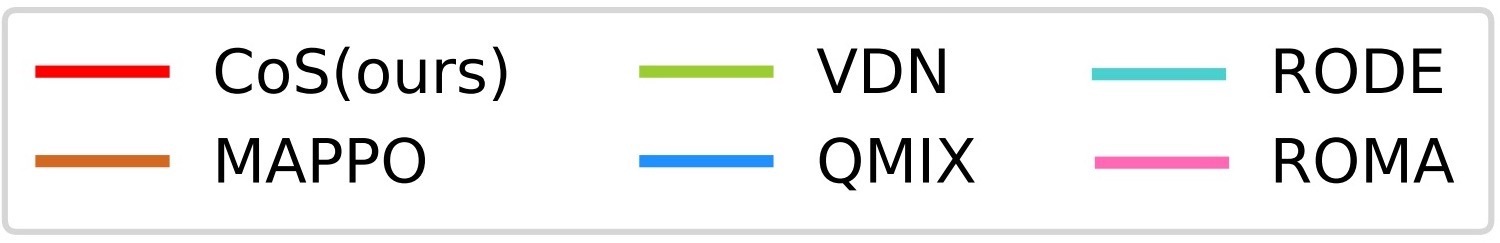}
		\vspace{-15pt}
		\end{minipage}
	}\vspace{-5pt}   
 
    \setcounter{subfigure}{0}
    \subfigure[\scriptsize 3vs1]{
    \label{fig:grf3-1}
		\begin{minipage}[t]{0.22\textwidth}
			\centering
			\includegraphics[width=\textwidth]{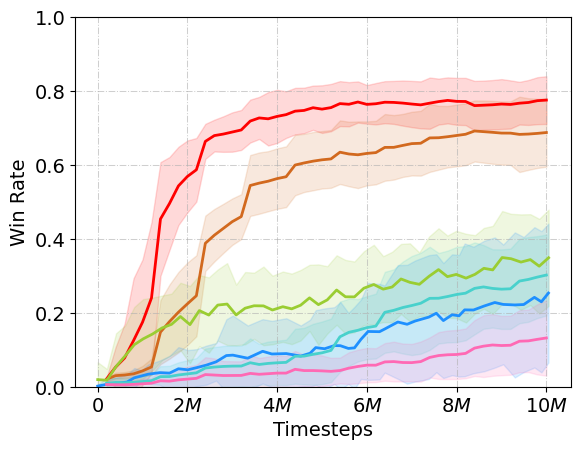}
		\end{minipage}
	}
	\subfigure[\scriptsize  c\_easy]{
 \label{fig:grf-cou-easy}
		\begin{minipage}[t]{0.22\textwidth}
			\centering
			\includegraphics[width=\textwidth]{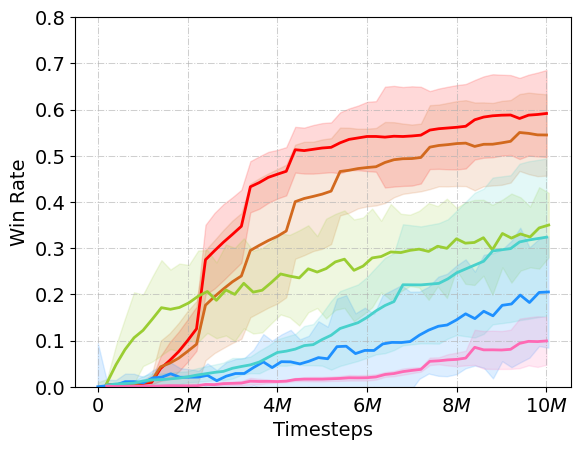}
		\end{minipage}
	}
	\centering
 \subfigure[\scriptsize c\_hard]{
    \label{fig:grf-cou-hard}
		\begin{minipage}[t]{0.22\textwidth}
			\centering
			\includegraphics[width=\textwidth]{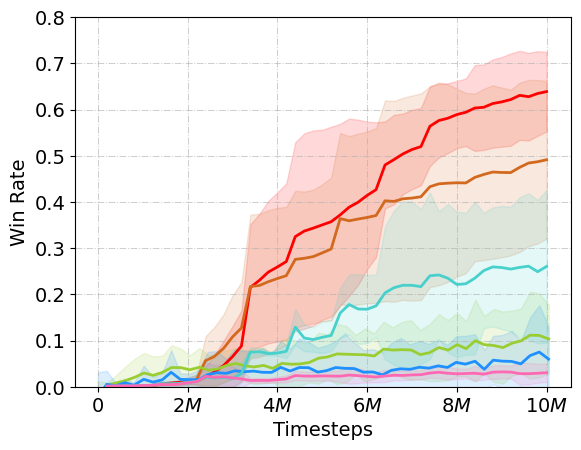}
		\end{minipage}
	}
	\subfigure[\scriptsize The statistic.]{
 \label{fig:grf-end}
		\begin{minipage}[t]{0.22\textwidth}
			\centering
			\includegraphics[width=\textwidth]{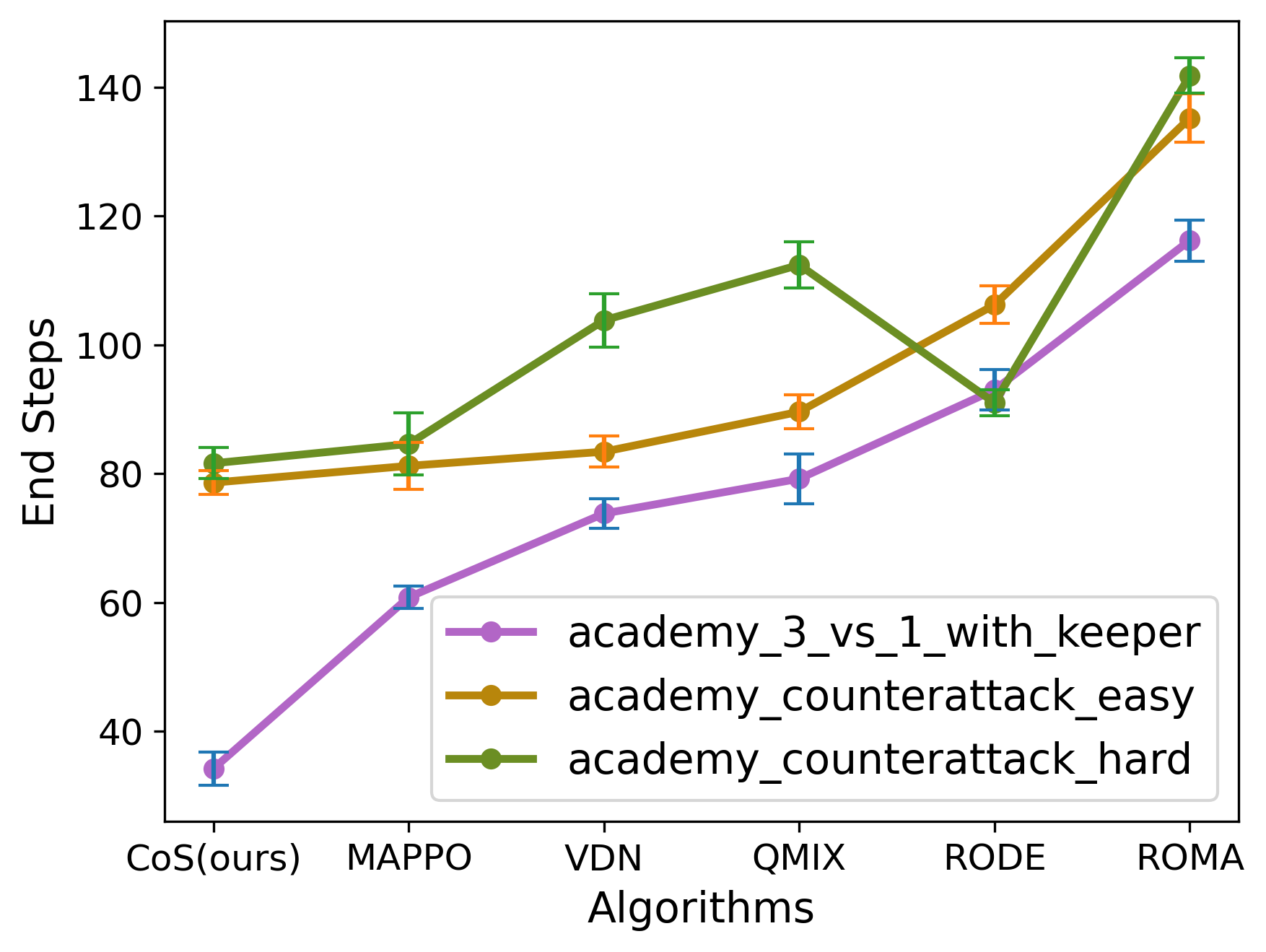}
		\end{minipage}
	}
	\centering
	\caption{The performance on three scenarios of GRF.}
 \vspace{-6pt}
	\label{fig:exp-grf}
\end{figure}

\vspace{-20pt}

\subsubsection{Performance on GRF.}

Further, we conduct experiments in google research football, a more dynamic and complicated benchmark to validate the effectiveness of our proposed CoS, shown in Figure~\ref{fig:exp-grf}.
Specifically, we choose three popular scenarios, including \textit{academy\_3\_vs\_1\_with\_keeper} (3vs1), \textit{academy\_counterattack\_easy} (c\_easy), and \textit{academy\_coun-terattack\_hard} (c\_hard).

In Figure~\ref{fig:exp-grf}(a-c), we observed that CoS consistently obtains higher performance than all the baselines in different scenarios of GRF, indicating that our method is robust and effective in complex and dynamic environments. Moreover, this performance improvement in GRF demonstrates that our approach excels at handling stochasticity, as the dynamic grouping strategy can extract more underlying details and make informed high-level decisions.  
Additionally, we report the average end steps of CoS in the three scenarios, shown in Figure~\ref{fig:grf-end}.
We test the trained model for 100 episodes and count the average end steps of every algorithm.
Specifically, we tested the trained model for 100 episodes and counted the average end steps of each algorithm. Notably, CoS completed the football game in the fewest steps compared to all the baselines, further validating the superiority of our proposed algorithm.


\subsection{Does the components of CoS Really Work?}

\begin{figure}[h!]
    \centering
    \setcounter{subfigure}{0}
    \subfigure[\scriptsize 4 agents on d-CN]{
    \label{fig:abl-4a}
		\begin{minipage}[t]{0.224\textwidth}
			\centering
			\includegraphics[width=\textwidth]{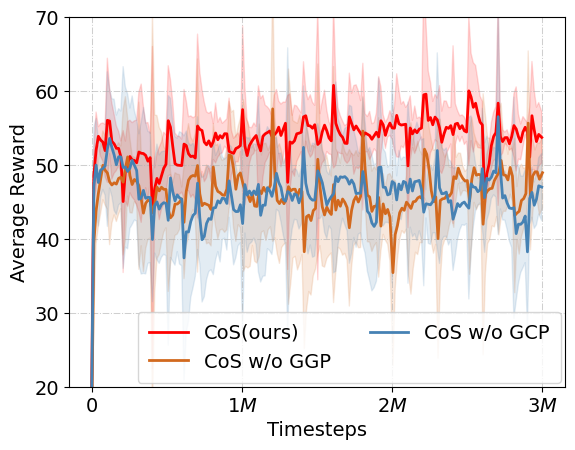}
		\end{minipage}
	}
	\subfigure[\scriptsize  6 agents on d-CN]{
 \label{fig:abl-6a}
		\begin{minipage}[t]{0.226\textwidth}
			\centering
			\includegraphics[width=\textwidth]{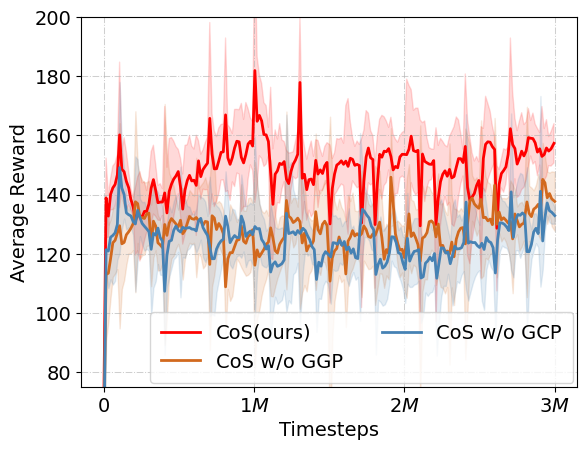}
		\end{minipage}
	}
	\centering
	\caption{The ablation study on d-CN of 4, 6 agents.}
	\label{fig:exp-abl}
\end{figure}

To evaluate the effectiveness of the components in CoS, we conduct the ablation study with the following configurations.
\begin{itemize}
    \item \textit{CoS(ours)}: The proposed framework.
    \item \textit{CoS w/o GGP}: Remove the group-guided policy.
    \item \textit{CoS w/o GCP}: Remove the group consensus policy.
\end{itemize}

As shown in Figure~\ref{fig:exp-abl}, the ablation study of 4 and 6 agents on d-CN shows the components in the CoS really work. The performance degradations of \textit{CoS w/o GGP} and \textit{CoS w/o GCP} show that the combination of these two modules is necessary, mainly due to two aspects. (1) GCP utilizes the extracted group consensus embeddings to make high-level decisions, which is conducive to long-term utility. (2) GGP perceives the individual observation from the true world, which can exploit underlying dynamics to make accurate decisions. Therefore, the fine-grained combination of these two modules
can achieve forward-looking guidance to specialize in a certain subtask and rational individual decision-making. 


\subsection{How Well Does the Group Consensus Embedding Perform?}

To investigate the discriminative power of the CoS model's group consensus embeddings, we visualize the embeddings of several scenarios on two tasks collected in the later stages of the training process. The goal is to show whether CoS can distinguish different groups and achieve separate consensus among every group, aiding in the collaboration of intra-groups in the decision-making process.

\begin{figure}[!ht]
 \vspace{-6pt}
    \centering
    \setcounter{subfigure}{0}
    \subfigure[\scriptsize 6 agents of c-CN]{
    \label{fig:c-CN-vis1-1}
		\begin{minipage}[t]{0.22\textwidth}
			\centering
			\includegraphics[width=\textwidth]{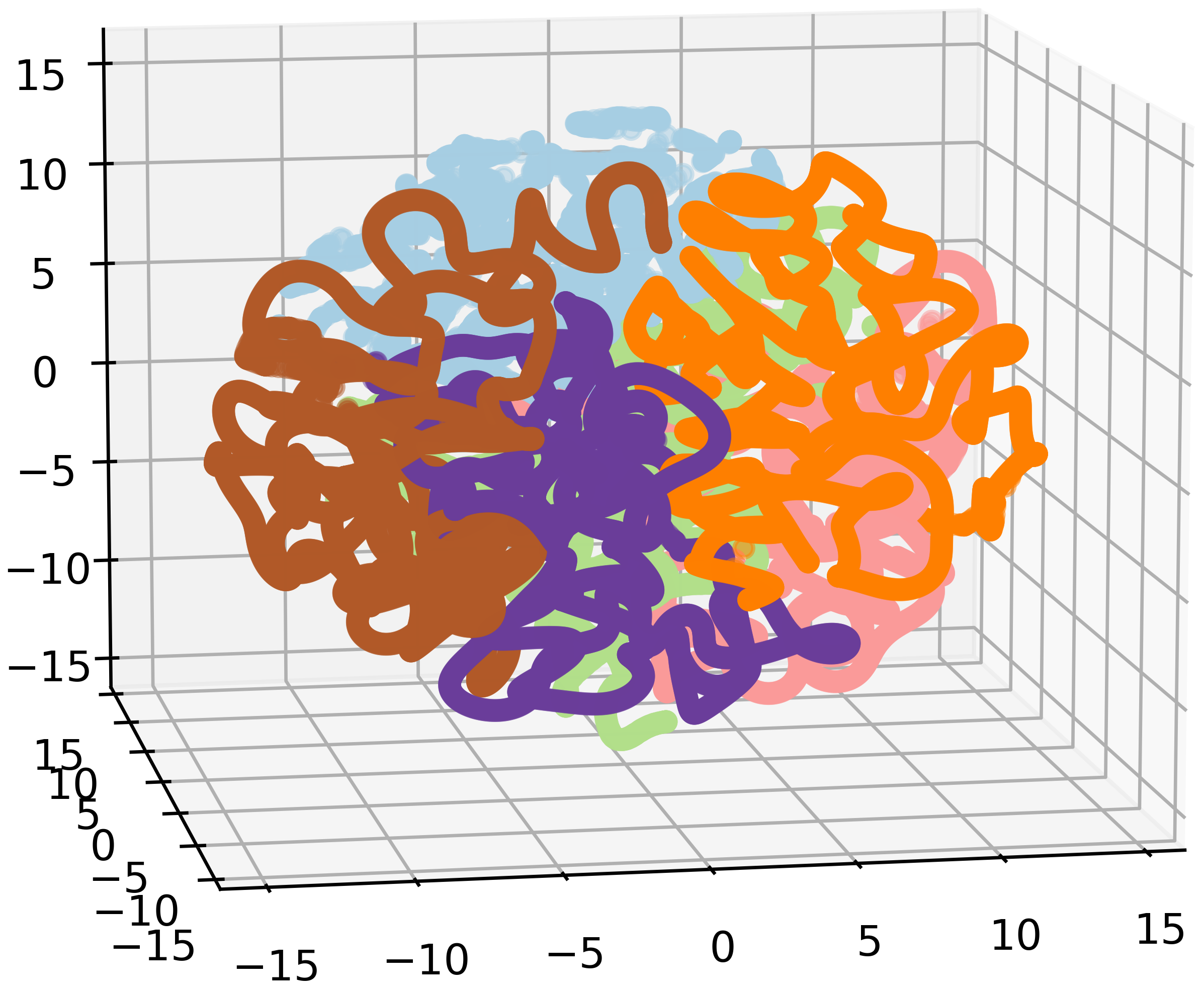}
		\end{minipage}
	}
	\subfigure[\scriptsize  10 agents of c-CN]{
 \label{fig:c-CN-vis1-2}
		\begin{minipage}[t]{0.22\textwidth}
			\centering
			\includegraphics[width=\textwidth]{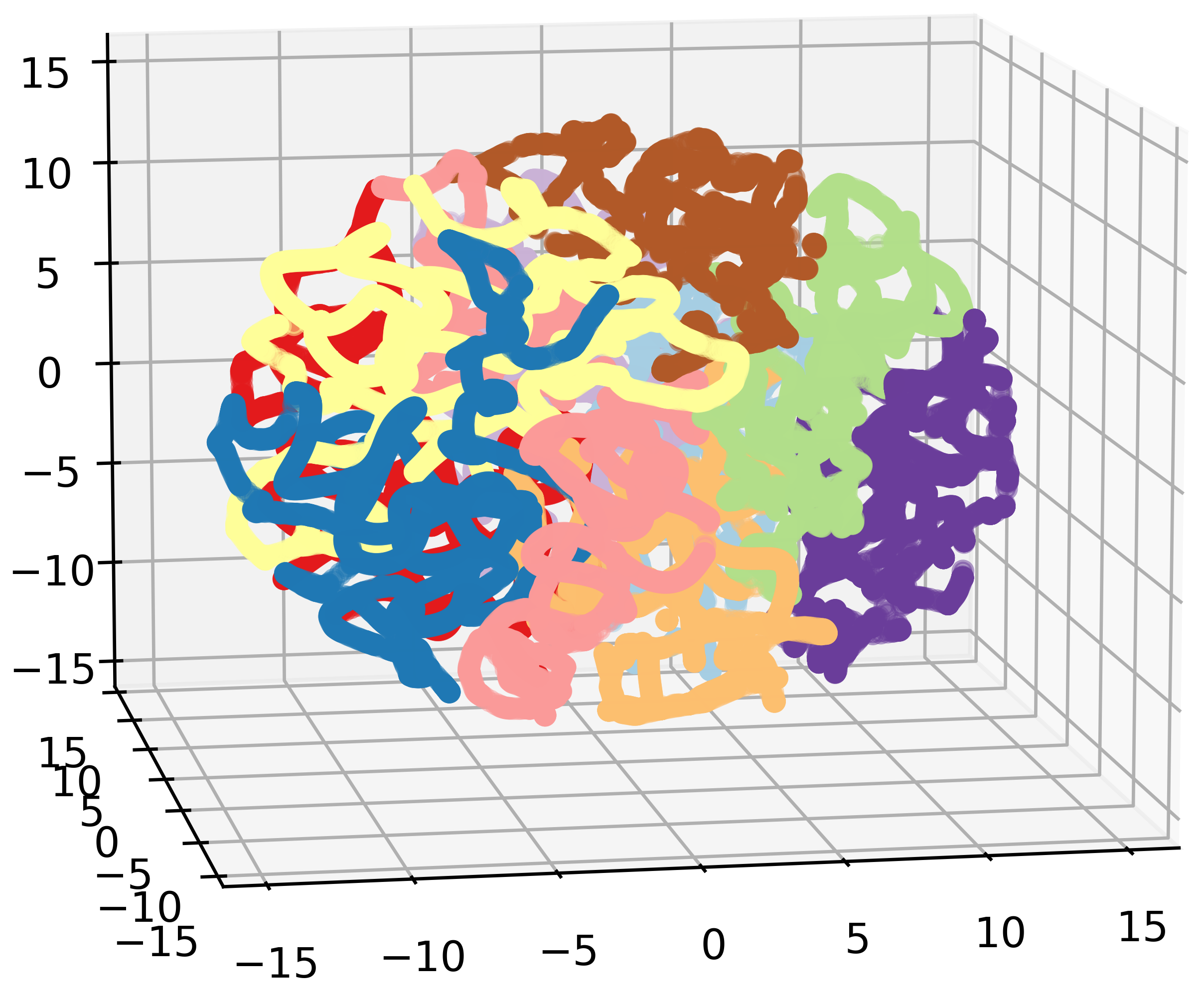}
		\end{minipage}
	}
 
    \subfigure[\scriptsize 3vs1 of c-CN]{
    \label{fig:grf-vis1-3}
		\begin{minipage}[t]{0.22\textwidth}
			\centering
			\includegraphics[width=\textwidth]{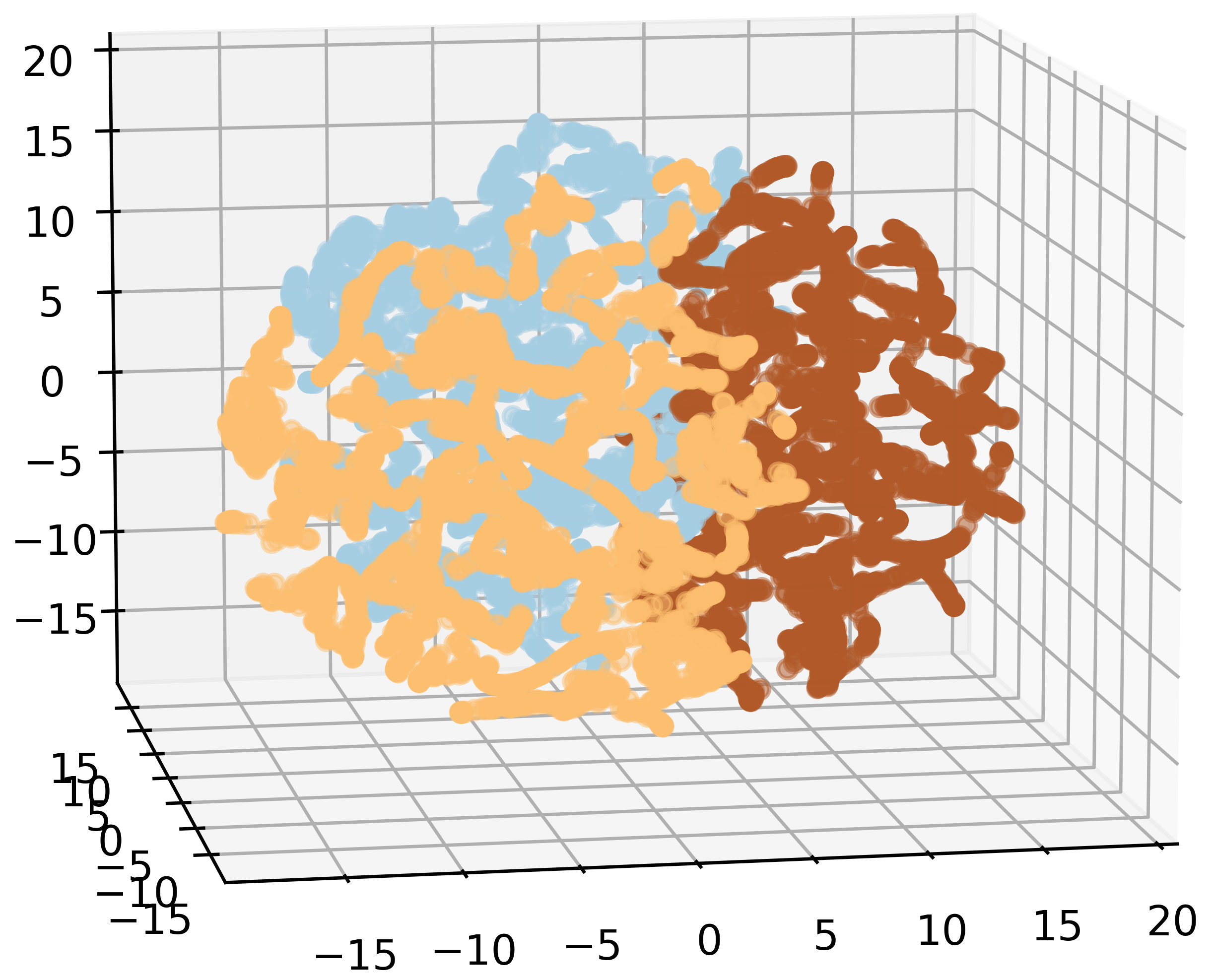}
		\end{minipage}
	}
	\subfigure[\scriptsize  c\_easy of GRF]{
 \label{fig:grf-vis1-4}
		\begin{minipage}[t]{0.22\textwidth}
			\centering
			\includegraphics[width=\textwidth]{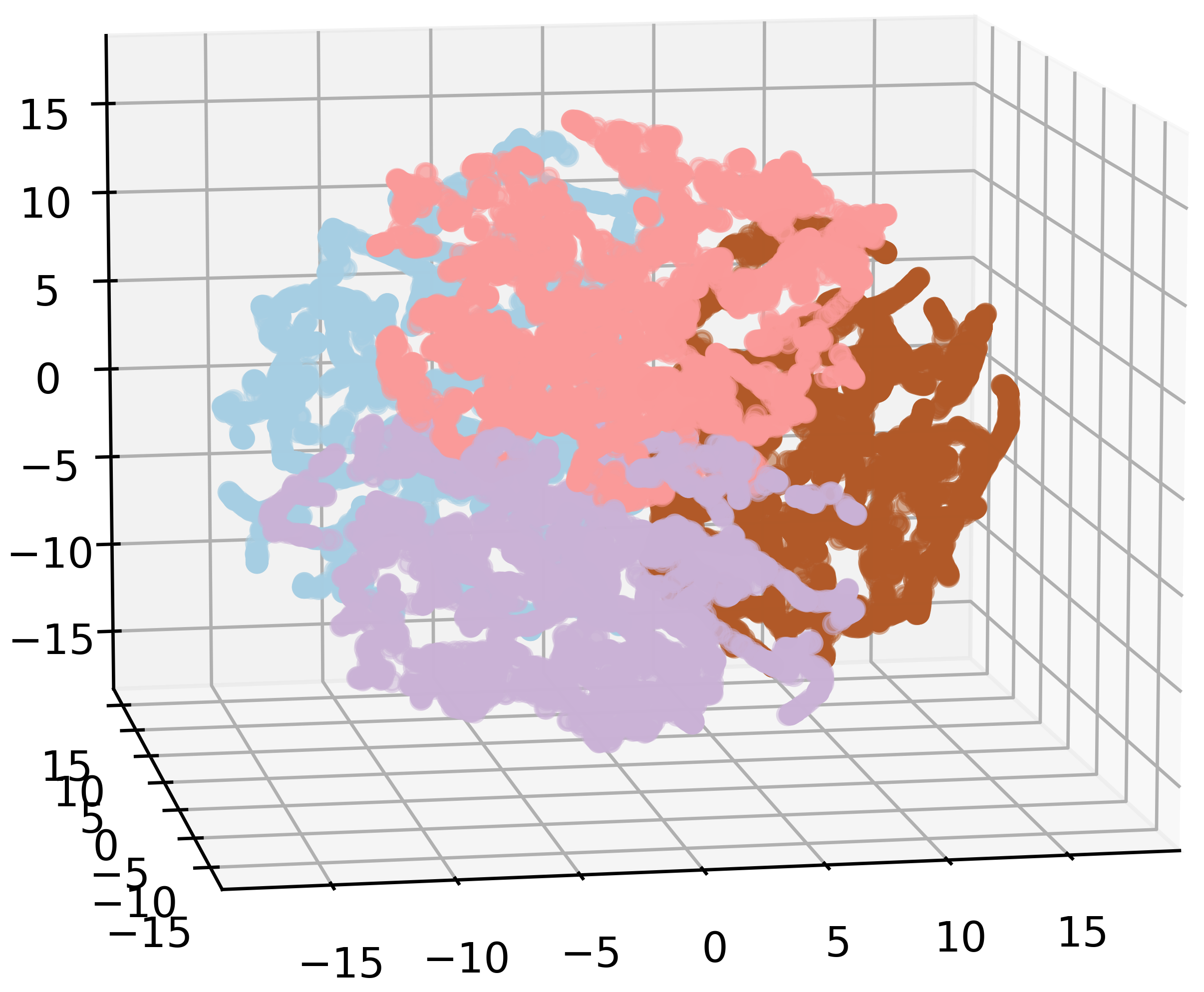}
		\end{minipage}
	}
 
	\centering
	\caption{Visualization of the group consensus embeddings on c-CN and GRF. We sample 5000 points for each group.}
	\label{fig:exp-vis1}
\end{figure}

Due to the assumption that $K$ agents can group into $K$ groups maximally, the number of group clusters is the same as the number of agents in each scenario. As shown in Figure~\ref{fig:exp-vis1}, the 3D visualizations demonstrate that CoS can consistently produce distinct group embeddings in all the scenarios. 
We showcase the visualizations obtained by applying the T-SNE technique to the group consensus embeddings saved from the last 5000 training steps.
Each cluster in the plots represents a specific group. Each point in the cluster corresponds to a vector in the group codebook. The distinguishable border highlights the ability of CoS to learn consistent group consensus representations.

Overall, these results demonstrate that our proposed method can facilitate intra-group collaboration and decision-making by producing effective group consensus embeddings that can discriminate between different groups in the decision-making process.



 \vspace{-16pt}

\subsection{Has CoS Learned to Group?}

\begin{figure}[!ht]
 \vspace{-6pt}
    \centering
    \setcounter{subfigure}{0}
    \subfigure[\scriptsize an example of d-CN]{
    \label{fig:d-CN-vis2-1}
		\begin{minipage}[t]{0.22\textwidth}
			\centering
			\includegraphics[width=\textwidth]{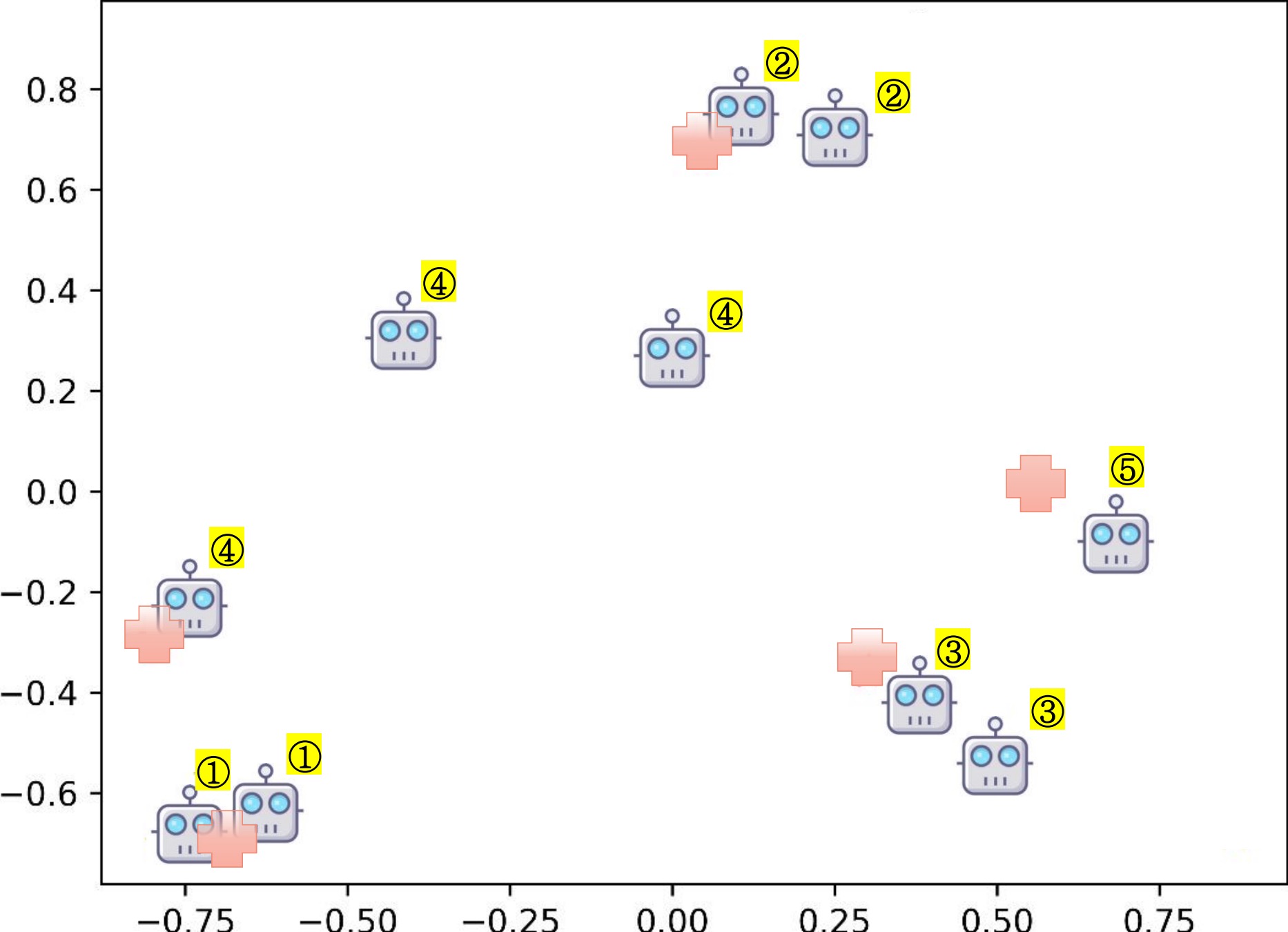}
		\end{minipage}
	}
	\subfigure[\scriptsize  group number vs. reward]{
 \label{fig:d-CN-vis2-2}
		\begin{minipage}[t]{0.22\textwidth}
			\centering
			\includegraphics[width=\textwidth]{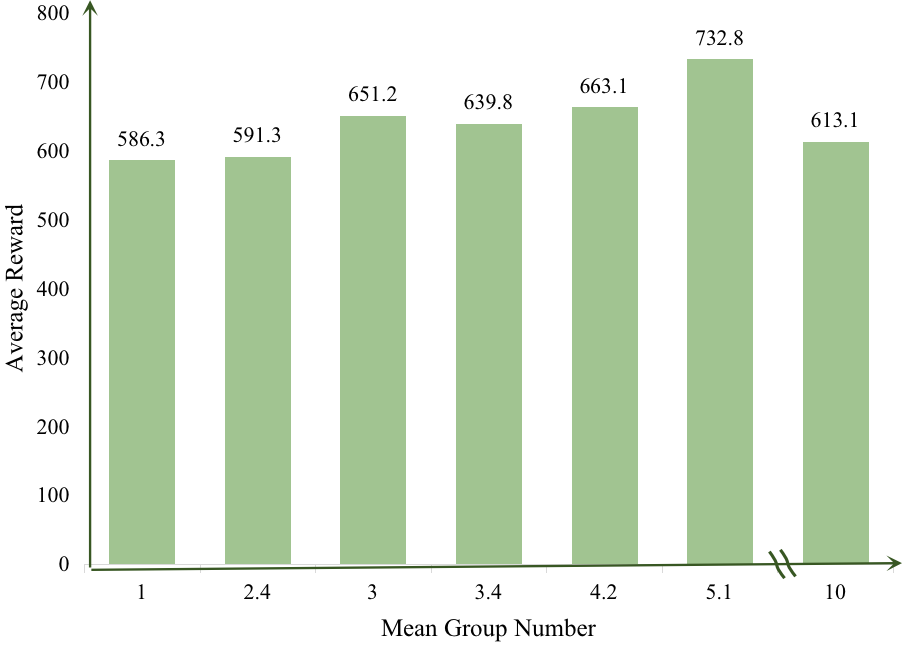}
		\end{minipage}
	}
	\centering
	\caption{Visualization of Grouping Effects.}
	\label{fig:exp-vis2}
\end{figure}

To assess CoS's ability to group and its impact on the collaborative behavior of agents, we conducted some visualizations and analyses. First, at time step $t=20$ during one testing episode, we visualize the grouping effect, as shown in Figure~\ref{fig:d-CN-vis2-1}, which illustrates that agents grouped by the same number in yellow exhibit similar behavior, indicating that CoS has learned to divide the agents into groups and promote collaboration.

To further examine the effect of grouping on performance, we evaluate the average rewards of 100 episodes at different checkpoints with varying numbers of groups, as shown in Figure~\ref{fig:d-CN-vis2-2}. Our results show that in the scenario with 10 agents, dividing them into 5 groups achieves the highest utility, consistent with the environmental setting. This finding highlights the importance of an effective grouping mechanism and demonstrates the capability of CoS to optimize the group sizes based on the task requirements.


 \vspace{-16pt}

\section{Conclusions}

In this paper, we propose CoS, a novel consensus-oriented strategy to promote multi-agent collaboration. First, CoS leverages the vector quantized variational autoencoder to extract the distinguishable and stable group consensus embeddings. Furthermore, the embeddings are used to assist global and individual decisions through the proposed group consensus policy and group-guided policy, respectively. Our empirical results on three benchmarks show that CoS significantly outperforms existing state-of-the-art methods. Additionally, the ablation study demonstrates the necessity of combining the two policies, and the visualizations further validate the effectiveness of CoS in promoting collaboration among agents. In the future, it is an interesting direction to investigate universal group embeddings across different tasks.

\clearpage

\bibliography{ecai}

\begin{thebibliography}{10}

\bibitem{agarwal2021contrastive}
Rishabh Agarwal, Marlos~C Machado, Pablo~Samuel Castro, and Marc~G Bellemare,
  `Contrastive behavioral similarity embeddings for generalization in
  reinforcement learning', {\em arXiv preprint arXiv:2101.05265}, (2021).

\bibitem{alam2016multi}
Md~Golam~Rabiul Alam, Yan~Kyaw Tun, and Choong~Seon Hong, `Multi-agent and
  reinforcement learning based code offloading in mobile fog', in {\em 2016
  International Conference on Information Networking}, pp. 285--290, (2016).

\bibitem{aman2021knowledge}
Bogdan Aman and Gabriel Ciobanu, `Knowledge dynamics and behavioural
  equivalences in multi-agent systems', {\em Mathematics}, {\bf 9}(22),  2869,
  (2021).

\bibitem{das2019tarmac}
Abhishek Das, Th{\'e}ophile Gervet, Joshua Romoff, Dhruv Batra, Devi Parikh,
  Mike Rabbat, and Joelle Pineau, `Tarmac: Targeted multi-agent communication',
  in {\em International Conference on Machine Learning}, pp. 1538--1546,
  (2019).

\bibitem{du2021flowcomm}
Yali Du, Bo~Liu, Vincent Moens, Ziqi Liu, Zhicheng Ren, Jun Wang, Xu~Chen, and
  Haifeng Zhang, `Learning correlated communication topology in multi-agent
  reinforcement learning', {\em International Conference on Autonomous Agents
  and Multi-Agent Systems}, (2021).

\bibitem{fan2021multi}
Yue Fan and Xiuli Ma, `Multi-vector embedding on networks with taxonomies',
  (2021).

\bibitem{ferber1998meta}
Jacques Ferber and Olivier Gutknecht, `A meta-model for the analysis and design
  of organizations in multi-agent systems', in {\em Proceedings international
  conference on multi agent systems (Cat. No. 98EX160)}, pp. 128--135, (1998).

\bibitem{ferber2004agents}
Jacques Ferber, Olivier Gutknecht, and Fabien Michel, `From agents to
  organizations: an organizational view of multi-agent systems', in {\em
  International workshop on agent-oriented software engineering}, pp. 214--230,
  (2004).

\bibitem{fervari2022bisimulations}
Raul Fervari, Fernando~R Vel{\'a}zquez-Quesada, and Yanjing Wang,
  `Bisimulations for knowing how logics', {\em The Review of Symbolic Logic},
  {\bf 15}(2),  450--486, (2022).

\bibitem{fu2022lilac}
Yuqian Fu, Jiajun Chai, Yuanheng Zhu, and Dongbin Zhao, `Lilac: Learning a
  leader for cooperative reinforcement learning', in {\em 2022 IEEE Conference
  on Games (CoG)}, pp. 49--55, (2022).

\bibitem{fujimoto2019off}
Scott Fujimoto, David Meger, and Doina Precup, `Off-policy deep reinforcement
  learning without exploration', in {\em International conference on machine
  learning}, pp. 2052--2062, (2019).

\bibitem{ha2016hypernetworks}
David Ha, Andrew Dai, and Quoc~V Le, `Hypernetworks', {\em arXiv preprint
  arXiv:1609.09106}, (2016).

\bibitem{hu2022policy}
Siyi Hu, Chuanlong Xie, Xiaodan Liang, and Xiaojun Chang, `Policy diagnosis via
  measuring role diversity in cooperative multi-agent rl', in {\em
  International Conference on Machine Learning}, pp. 9041--9071, (2022).

\bibitem{iqbal2021randomized}
Shariq Iqbal, Christian A~Schroeder De~Witt, Bei Peng, Wendelin B{\"o}hmer,
  Shimon Whiteson, and Fei Sha, `Randomized entity-wise factorization for
  multi-agent reinforcement learning', in {\em International Conference on
  Machine Learning}, pp. 4596--4606, (2021).

\bibitem{jiang2018graph}
Jiechuan Jiang, Chen Dun, Tiejun Huang, and Zongqing Lu, `Graph convolutional
  reinforcement learning', {\em arXiv preprint arXiv:1810.09202}, (2018).

\bibitem{jiang2018atoc}
Jiechuan Jiang and Zongqing Lu, `Learning attentional communication for
  multi-agent cooperation', {\em Annual Conference on Neural Information
  Processing Systems}, (2018).

\bibitem{jiang2022multi}
Qize Jiang, Minhao Qin, Shengmin Shi, Weiwei Sun, and Baihua Zheng,
  `Multi-agent reinforcement learning for traffic signal control through
  universal communication method', {\em International Joint Conference on
  Artificial Intelligence}, (2022).

\bibitem{kurach2020google}
Karol Kurach, Anton Raichuk, Piotr Sta{n}czyk, Micha{l} Zaj{k{a}}c, Olivier
  Bachem, Lasse Espeholt, Carlos Riquelme, Damien Vincent, Marcin Michalski,
  Olivier Bousquet, et~al., `Google research football: A novel reinforcement
  learning environment', in {\em Proceedings of the AAAI Conference on
  Artificial Intelligence}, volume~34, pp. 4501--4510, (2020).

\bibitem{lhaksmana2013role}
Kemas~M Lhaksmana, Yohei Murakami, and Toru Ishida, `Role modeling for adaptive
  multiagent systems engineering', in {\em 2013 IEEE/WIC/ACM International
  Joint Conferences on Web Intelligence (WI) and Intelligent Agent Technologies
  (IAT)}, volume~2, pp. 287--292, (2013).

\bibitem{li2022ace}
Chuming Li, Jie Liu, Yinmin Zhang, Yuhong Wei, Yazhe Niu, Yaodong Yang, Yu~Liu,
  and Wanli Ouyang, `Ace: Cooperative multi-agent q-learning with bidirectional
  action-dependency', {\em arXiv preprint arXiv:2211.16068}, (2022).

\bibitem{liu2022self}
Kai Liu, Yuyang Zhao, Gang Wang, and Bei Peng, `Self-attention-based
  multi-agent continuous control method in cooperative environments', {\em
  Information Sciences}, {\bf 585},  454--470, (2022).

\bibitem{liu2020when2com}
Yen-Cheng Liu, Junjiao Tian, Nathaniel Glaser, and Zsolt Kira, `When2com:
  Multi-agent perception via communication graph grouping', in {\em Proceedings
  of the IEEE/CVF Conference on computer vision and pattern recognition}, pp.
  4106--4115, (2020).

\bibitem{liu2022rogc}
Yuntao Liu, Yuan Li, Xinhai Xu, Donghong Liu, and Yong Dou, `Rogc:
  Role-oriented graph convolution based multi-agent reinforcement learning', in
  {\em 2022 IEEE International Conference on Multimedia and Expo (ICME)}, pp.
  1--6, (2022).

\bibitem{lowe2017multi}
Ryan Lowe, Yi~I Wu, Aviv Tamar, Jean Harb, OpenAI Pieter~Abbeel, and Igor
  Mordatch, `Multi-agent actor-critic for mixed cooperative-competitive
  environments', {\em Advances in neural information processing systems}, {\bf
  30}, (2017).

\bibitem{nickel2017poincare}
Maximillian Nickel and Douwe Kiela, `Poincar{\'e} embeddings for learning
  hierarchical representations', {\em Advances in neural information processing
  systems}, {\bf 30}, (2017).

\bibitem{niu2021multi}
Yaru Niu, Rohan~R Paleja, and Matthew~C Gombolay, `Multi-agent graph-attention
  communication and teaming.', in {\em AAMAS}, pp. 964--973, (2021).

\bibitem{odell2002role}
James~J Odell, H~Van Dyke~Parunak, and Mitchell Fleischer, `The role of roles
  in designing effective agent organizations', in {\em International Workshop
  on Software Engineering for Large-Scale Multi-agent Systems}, pp. 27--38,
  (2002).

\bibitem{pavon2003agent}
Juan Pav{\'o}n and Jorge G{\'o}mez-Sanz, `Agent oriented software engineering
  with ingenias', in {\em International Central and Eastern European Conference
  on Multi-Agent Systems}, pp. 394--403, (2003).

\bibitem{phan2021vast}
Thomy Phan, Fabian Ritz, Lenz Belzner, Philipp Altmann, Thomas Gabor, and
  Claudia Linnhoff-Popien, `Vast: Value function factorization with variable
  agent sub-teams', {\em Advances in Neural Information Processing Systems},
  {\bf 34},  24018--24032, (2021).

\bibitem{poincare1882theorie}
Henri Poincar{\'e}, `Th{\'e}orie des groupes fuchsiens', {\em Acta
  mathematica}, {\bf 1}(1),  1--62, (1882).

\bibitem{tianhypertron}
Yu~Tian, Xingliang Huang, Ruigang Niu, Hongfeng Yu, Peijin Wang, and Xian Sun,
  `Hypertron: Explicit social-temporal hypergraph framework for multi-agent
  forecasting', (2022).

\bibitem{uchendu2022jump}
Ikechukwu Uchendu, Ted Xiao, Yao Lu, Banghua Zhu, Mengyuan Yan, Jos{\'e}phine
  Simon, Matthew Bennice, Chuyuan Fu, Cong Ma, Jiantao Jiao, et~al.,
  `Jump-start reinforcement learning', {\em arXiv preprint arXiv:2204.02372},
  (2022).

\bibitem{van2017vqvae}
Aaron Van Den~Oord, Oriol Vinyals, et~al., `Neural discrete representation
  learning', {\em Advances in neural information processing systems}, {\bf 30},
  (2017).

\bibitem{wang2021rode}
T~Wang, T~Gupta, B~Peng, A~Mahajan, S~Whiteson, and C~Zhang, `Rode: learning
  roles to decompose multi- agent tasks', in {\em Proceedings of the
  International Conference on Learning Representations}, (2021).

\bibitem{wang2020roma}
Tonghan Wang, Heng Dong, Victor Lesser, and Chongjie Zhang, `Roma: Multi-agent
  reinforcement learning with emergent roles', in {\em International Conference
  on Machine Learning}, pp. 9876--9886, (2020).

\bibitem{wu2020multi}
Tong Wu, Pan Zhou, Kai Liu, Yali Yuan, Xiumin Wang, Huawei Huang, and
  Dapeng~Oliver Wu, `Multi-agent deep reinforcement learning for urban traffic
  light control in vehicular networks', {\em IEEE Transactions on Vehicular
  Technology}, {\bf 69}(8),  8243--8256, (2020).

\bibitem{xu2022consensus}
Zhiwei Xu, Bin Zhang, Dapeng Li, Zeren Zhang, Guangchong Zhou, and Guoliang
  Fan, `Consensus learning for cooperative multi-agent reinforcement learning',
  {\em Advances in Neural Information Processing Systems}, (2022).

\bibitem{yang2022ldsa}
Mingyu Yang, Jian Zhao, Xunhan Hu, Wengang Zhou, and Houqiang Li, `Ldsa:
  Learning dynamic subtask assignment in cooperative multi-agent reinforcement
  learning', {\em arXiv preprint arXiv:2205.02561}, (2022).

\bibitem{yuan2022multi}
Lei Yuan, Chenghe Wang, Jianhao Wang, Fuxiang Zhang, Feng Chen, Cong Guan,
  Zongzhang Zhang, Chongjie Zhang, and Yang Yu, `Multi-agent concentrative
  coordination with decentralized task representation', in {\em International
  Joint Conference on Artificial Intelligence}, (2022).

\bibitem{zhang2011coordinated}
Chongjie Zhang and Victor Lesser, `Coordinated multi-agent reinforcement
  learning in networked distributed pomdps', in {\em Twenty-Fifth AAAI
  Conference on Artificial Intelligence}, (2011).

\bibitem{zhang2022common}
Xianjie Zhang, Yu~Liu, Hangyu Mao, and Chao Yu, `Common belief multi-agent
  reinforcement learning based on variational recurrent models', {\em
  Neurocomputing}, {\bf 513},  341--350, (2022).

\end{thebibliography}

\normalsize 
\newpage 
\onecolumn 
\setlength{\baselineskip}{20pt} 

\appendix
\setcounter{tocdepth}{2}
\onecolumn
{\centering\section{Appendix}}
\label{sec:app}


\subsection{Network Architecture}
\label{app:net}

There is a summary of all the neural networks used in our framework about the network structure, layers, and activation functions.
\begin{table}[h!]
    \centering
    \begin{tabular}{ccccc}
       \toprule
       ~ & Network Structure &  Layers & Hidden Size & Activation Functions  \\
       \midrule
       History Encoder & CausalSelfAttention & 1 &  -  & ReLu \\
       History Decoder & MLP & 2 &  64  & ReLu   \\
       Group Codebook &   nn.Embedding & 1 & [N, 32] & None \\
       Feature Extractor & CNN+FiLMedBlock & - & - & ReLu \\
       Group-Guided Policy &   Hyper-MLP & 2 & 32 & Tanh \\
       Group Consensus Policy  &   MLP+RNN+MLP & 3 & [64]+[64]+[32] & Tanh \\
       Policy Critic & MLP & 2 & 64 & Tanh\\
       \bottomrule
    \end{tabular}
   \caption{The Summary for Network Architecture}
   \label{tab:net}
\end{table}

Please note that '$-$' denote that we refer readers to check the open source repository CausalSelfAttention, see \href{https://github.com/sachiel321/Efficient-Spatio-Temporal-Transformer}{https://github.com/sachiel321/Efficient-Spatio-Temporal-Transformer}.

\subsection{Parameter Settings}
\label{app:para}

There are our hyper-parameter settings for the training of Vector Quantized Group Consensus and Group Consensus-oriented Strategy, shown in Table~\ref{tab:para1} and Table~\ref{tab:para2}, respectively.

\vspace{10pt}

\begin{minipage}{\textwidth}
\begin{minipage}[t]{0.5\textwidth}

\centering
    \begin{tabular}{cc}
       \toprule
       Description &  Value  \\
       \midrule
       optimizer & $AdamW$  \\
       lr & $5*10^{-4}$  \\
        group embedding size  &   $32$ \\
        context length         &    $1$ \\
        model type        &     $state \ only$ \\
        attention head &    $4$ \\
        actor embedding size &    $32$ \\
        group codebook $K$ &    $agent \ number$ \\
        encoder coefficient $\beta$ &    $1$ \\
       \bottomrule
    \end{tabular}
\makeatletter\def\@captype{table}\makeatother\caption{Hyper-parameters of Vector Quantized Group Consensus}
\label{tab:para1}
\end{minipage}
\begin{minipage}[t]{0.5\textwidth}
\centering
    \begin{tabular}{cc}
       \toprule
       Description &  Value \\
       \midrule
       optimizer       &    Adam   \\
        $\alpha$        &    $10^{-4}$  \\
        $\beta_1$       &    $0.9$  \\
        $\beta_2$       &    $0.999$  \\
        $\varepsilon$-greedy $\varepsilon$   &    $10^{-5}$  \\
        clipping $\epsilon$       &   $0.2$ \\
        seed             &   $[0,10)$  \\
        number of process &  $64$  \\
        lr             &     $10^{-4}$  \\
        eval interval  &     $400000$  \\
        eval episodes  &     $100$   \\
        jump & $0.3$ \\
       \bottomrule
    \end{tabular}

\makeatletter\def\@captype{table}\makeatother\caption{Hyper-parameters of Group Consensus-oriented Strategy}
\label{tab:para2}
\end{minipage}
\end{minipage}

By the way, 
we refer readers to the source code in the supplementary to check the detailed hyper-parameters.

\subsection{The Pseudo-code of CoS}
\label{app:algo}

The main procedures of our proposed CoS are summarized as Algorithm~\ref{algo:cos}.

\begin{algorithm}[ht!]
\caption{The pseudo-code of CoS.}
\label{algo:cos}
\textbf{Ensure} vector quantized group consensus $\rho=\{q, \{e^j\}_{j=1}^K, p\}$;\\
\textbf{Ensure} group consensus policy $\pi^{gc^i}$, group-guided policy $\pi^{gg^i}$ for agent $i$, and critic $Q_{\pi}$;\\
\textbf{Initialize}~the parameters $\theta$ 
for the vector quantized group consensus, the parameter $\phi^{gc^i}, \phi^{gg^i}$ for the group consensus policy and group-guided policy, where $i=1,...,N$, and
$\varphi$ for the critic network;\\
\textbf{Initialize}~jump\_rate=0.3;\\
\For {each episode}{
Initial state;\\
 \For{ each timestep}{
Compute history encoder output $z_e(\tau^i)=q_{\theta}(\cdot|\tau^i)$;\\
Quantize $z_e$ as $z_q$ with the group codebook $\bm{e}$;\\
Obtain the group consensus embedding $e^j$;\\
Reconstruct the input $\hat \tau^i = p_{\theta}(\cdot|z^q)$;\\
Compute the group-guided action $a^{{gg}^i_j}=\pi^{gg^i}(\cdot | o^i, e^j)$;\\
Obtain action for agent $i$ as $u^i=a^{gg^i_j}$;\\
\If{ $\#episodes > jump\_rate*total\_number$}{
Compute the group consensus action $a^{gc^i_j} = \pi^{gc^i}(\cdot|e^j)$;\\
Obtain action for the agent $i$ as $u^i=a^{gc^i_j}+a^{gg^i_j}$; \\
// Note: the sum of actions for continuous action space; the sum of action distributions for discrete action space;
}
Obtain the joint action $\bm{u}=\{u^1,u^2,...,u^N\}$;\\
Receive reward $r_t$ and observe next state $\{o_t^i\}_{i=1}^N$;\\
}
// Training \\
Update the group consensus policy with Equation~(\ref{eq:gcp});\\ 
Update the group-guided policy with Equation~(\ref{eq:ggp});\\ 
Update the vector quantized group consensus with Equation~(\ref{eq:vqgc});\\
Update the critic network with TD error.
}

\end{algorithm}

\subsection{The Detailed Description of GRF}
\label{app:grf}

\paragraph{\textbf{Observations}}
The environment exposes the raw observations as Table~\ref{tab:grf-info}.
We use the \textit{Simple115StateWrapper}\footnote{We refer the reader to:\url{https://github.com/google-research/football} for details of encoded information.} as the simplified representation of a game state encoded with 115 floats.

\begin{sidewaystable}[!ht]
    \centering
    \caption{The detailed descriptions of the information included in the raw observation of GRF.}
\label{tab:grf-info}
    \begin{tabular}{cc|cc}
    \toprule
        \textbf{Information} & \textbf{Descriptions} & \textbf{Information} & \textbf{Descriptions} \\ \hline
        ~ & position of ball & ~ & ~ \\ 
        ~ & direction of ball & ~ & controlled player index \\ 
        Ball & rotation angles of ball & Controlled player & designated player index \\ 
        ~ & owned team of ball & ~ & active action \\ 
        ~ & owned player of ball & ~ & ~ \\ \hline
        ~ & position of players in left team & ~ & position of players in right team \\ 
        ~ & direction of players in left team & ~ & direction of players in right team \\ 
        Left Team & tiredness factor of players & Right Team & tiredness factor of players \\ 
        ~ & numbers of players with yellow card & ~ & numbers of players with yellow card \\ 
        ~ & whether a player got a red card & ~ & whether a player got a red card \\ 
        ~ & roles of players & ~ & roles of player \\ \hline
        ~ & goals of left and right teams & ~ & ~ \\ 
        Match state & left steps & Screen & rendered screen \\ 
        ~ & current game mode & ~ & ~ \\ \bottomrule
    \end{tabular}
\end{sidewaystable}

\paragraph{\textbf{Actions}}
The number of actions available to an individual agent can be denoted as $|\mathcal{A}| = 19$.
The standard move actions (in $8$ directions) include $\mathcal{A}_{move} = \{Top, Bottom, Left, Right,Top-Left, Top-Right, Bottom-Left, Bottom-Right\}$.
Moreover, the actions represent different ways to kick the ball is 
$\mathcal{A}_{kick} = \{Short Pass, High Pass,Long Pass, Shot,Do-Nothing,Sliding,Dribble,Stop-Dribble,Sprint, Stop-Moving, Stop-Sprint\}$.

\paragraph{\textbf{Rewards}}
The reward function mainly includes two parts. The first is $SCORING$, which corresponds to the natural reward where the team obtains $+1$ when scoring a goal and $-1$ when losing one to the opposing team. The second part is $CHECKPOINT$, which is proposed to address the issue of sparse rewards. It is encoded with  domain
knowledge by an additional auxiliary reward contribution.

\paragraph{\textbf{Scenarios}}
We conduct the experiments on the following scenarios.
\begin{itemize}
    \item academy\_3\_vs\_1\_with\_keeper: Three of our players try to score from the edge of the box, one on each side, and the other at the center. Initially, the player at the center has the ball and is facing the defender. There is an opponent keeper.
    \item academy\_counterattack\_easy: 4 versus 1 counter-attack with keeper; all the remaining players of both teams run back towards the ball.
    \item academy\_counterattack\_hard: 4 versus 2 counter-attack with keeper; all the remaining players of both teams run back towards the ball. 
\end{itemize}

\end{document}